\documentclass[12pt]{iopart}
\input epsf
\usepackage{amsfonts,amssymb}
\usepackage{graphicx} 

\begin{document}

%\preprint{ }
\title[The frustration-based approach of supercooled liquids and the glass
transition]{The frustration-based approach of supercooled liquids and the glass
transition: a review and critical assessment}
\author{G. Tarjus$^1$, S. A. Kivelson$^2$, Z. Nussinov$^3$, P. Viot $^1$}
%\affiliation
\address
{$^1$ Laboratoire de Physique  Th\' eorique de la Mati\` ere Condens\' ee, Universit\' e
Pierre et
Marie Curie, 4 Place Jussieu, 75252 Paris Cedex 05, France}
%
%\affiliation
\address
{$^2$GLAM, Department of Physics, Stanford University, Stanford, CA 94305, USA}
%
%\affiliation
\address
{$^3$Theoretical Division, Los Alamos National Laboratory, Los Alamos, NM 87545,
and Department of Physics, Washington University, St Louis, MO 63160, USA}
\bibliographystyle{prsty}

\date{\today}
\begin{abstract}
One of the most spectacular phenomena in physics in terms of dynamical
range is the glass transition and  the associated slowing down of flow
and relaxation with decreasing temperature.  That it occurs in  many
different liquids seems  to call  for a  "universal" theory.  In  this
article, we review one such theoretical approach which is based on the
concept of "frustration".   Frustration  in this context describes  an
incompatibility between extension of the  locally preferred order in a
liquid and tiling of the whole space. We provide a critical assessment
of what has  been achieved  within this  approach and we  discuss  the
relation with other theories of the glass transition.
\end{abstract}
%\maketitle

%\vspace{1cm}
%\hrulefill\
%\narrowtext

%\def\be{\begin{equation}}
%\def\ee{\end{equation}}
%\def\bea{\begin{eqnarray}}
%\def\eea{\end{eqnarray}}
%\def\bit{\begin{itemize}}
%\def\eit{\end{itemize}}

\section{Introduction}\label{sec:introduction}

One of the disappointing aspects of  the study of the glass transition
is  that after  many  years  of  scholarly  effort there remains  wide
divergence  in   the   basic ways  in    which  relevant phenomena are
envisaged. By "basic ways to envisage phenomena"  we mean the physical
models upon  which relevant theories of structure  and dynamics can be
built, i.e., the    zeroth-order  descriptions which  with  but  minor
elaboration  should  enable      one   to  describe   universal     or
species-independent phenomena over a  wide range of control  variables
such as temperature and  pressure.  Perhaps such great differences  of
viewpoints    exist because no   such  zeroth-order description can be
constructed, but we do not believe this to be  the case. What might be
described as the "anomalous  slowing down" of  relaxation and flow  in
liquids  with the  approach to the   glass transition appears to  be a
sufficiently general phenomenon   to be describable by   a "universal"
theory.

We review in this article one such theoretical approach which has been
developed over the past two decades: it is based  on the idea that the
physical mechanism  which is  responsible  for glass  formation is the
ubiquitous        presence   of      "\textit{frustration}"         in
liquids.  "Frustration" in  this context describes  an incompatibility
between extension of the local order preferred  in a liquid and tiling
of the whole space.

After briefly summarizing the phenomenology of glassforming liquids in
light  of     what    we think     to     be the     relevant  physics
(section~\ref{sec:phen-what-there}), we   introduce   the  concept  of
frustration.   We  first describe  what  is  probably the most studied
example,   that    of frustrated    icosahedral   order  in   metallic
glassformers, and we  summarize the curved-space approach of geometric
frustration       (section       \ref{sec:geom-frustr-simple}).     In
section~\ref{sec:stat-mech-frustr}, we  discuss    the     statistical
mechanical  description  of  frustration in   liquids, with  both  the
underlying assumptions and  the  possible candidates  for  providing a
minimal theoretical  model  at a coarse-grained,  mesoscopic level. We
next present  the evidence and the consequences  of a generic property
shared by  the  frustrated models  for the  glass transition,  that of
"avoided criticality"    (section~\ref{sec:avoided-criticality}).   In
section~\ref{sec:phen-scal-appr}, we review a phenomenological scaling
approach to supercooled liquids built about an avoided critical point,
the       frustration-limited      domain    theory,     and        in
section~\ref{sec:comp-simul-simple}  we    summarize the    results of
extensive  computer   simulation  studies   of   3-dimensional Coulomb
frustrated  lattice models.   The  following  section  is  devoted  to
discussing  additional  phenomena  brought about  by  frustration, the
emergence of a complex free-energy     landscape characterized by    a
proliferation  of long-lived metastable  states  and  the presence  of
topological excitations (defects)   whose effective  dynamics  may  be
strongly  constrained;   this allows us   to make  contact  with other
theoretical          approaches       of       glass         formation
(section~\ref{sec:conn-other-appr}).                          Finally,
section~\ref{sec:conclusion} presents some concluding remarks.

This article is intended to provide a  critical assessment of what has
been achieved within  the  frustration-based approach of  glassforming
liquids, discussing the limitations and the open questions.

\section{Phenomenology: what is there to be explained?}\label{sec:phen-what-there}

What is meant by the "glass transition"  is not always well specified,
but  usually it refers  to  the passage  from liquid-like behavior  to
behavior characteristic of amorphous solids. The "glass transition" so
defined is not a thermodynamic, nor even  a dynamic, transition; it is
a  point of dynamic arrest  arbitrarily  but narrowly specified by the
available experimental techniques for studying  flow or relaxation. We
focus here on  the approach to this glass  transition, when the liquid
or the polymer can still be considered at equilibrium (although it may
be  "supercooled" and  therefore  in metastable equilibrium, the  most
stable phase being a crystal).

\begin{figure}\begin{centering}
\resizebox{10cm}{!}{\includegraphics{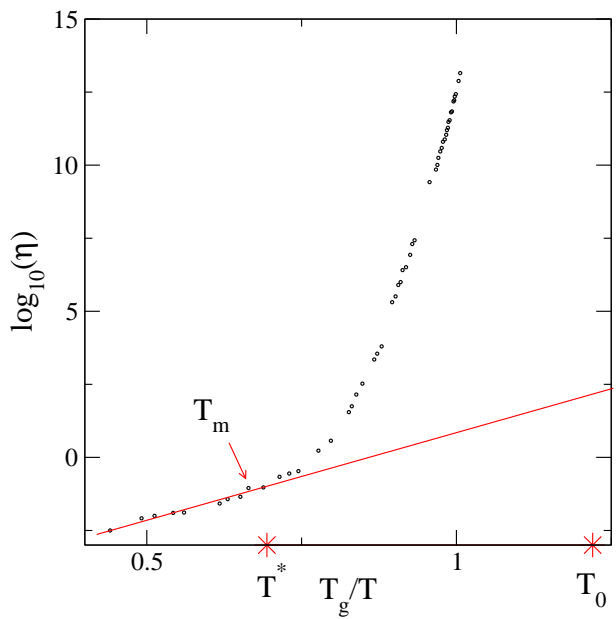}}
\caption{SuperArrhenius temperature dependence of the viscosity of liquid
ortho-terphenyl: logarithm (base $10$) of $\eta$ versus inverse temperature
$1/T$. The  straight line  is the extrapolation  of the  behavior in the
"ordinary" liquid range  (above $T_m$); whereas  the actual $T_g$  is around
$243K$, the  extrapolated  one is   at a  much lower temperature  around
$50-70K$. (Data taken from references cited in \cite{KTZK96}).} \label{fig:1}
\end{centering}\end{figure}

The distinctive property of  glassforming liquids and polymers  is the
dramatic  slowing  down  of  relaxation   and  flow  with   decreasing
temperature\cite{EAN96,DEBE01,TARKI01}.   The dynamical range of  this
phenomenon   is probably one  of   the most   spectacular in  physics:
viscosity and  main  ($\alpha$) relaxation  time may  increase   by some 15
orders     of     magnitude   for   a   mere     $30\%$     decrease in
temperature. Especially striking is  the  fact that for most   liquids
(notable  exceptions  are    network   forming   systems  with  strong
directional bonding such as  silica glasses), the slowing  down occurs
much more rapidly that one would have anticipated by extrapolating the
behavior of  the "ordinary" liquid, say  above its melting temperature
$T_{m}$.  For instance, extrapolation of the high-temperature data for
the viscosity of liquid ortho-terphenyl, which are well described by a
simple            Arrhenius           dependence,           $\eta \approx \eta_{\infty}
\exp\left[E_{\infty}/(k_{B}T)\right]$,  would  imply    a  glass  transition
temperature  (conventionally    defined as the   point   at  which the
viscosity takes a given  value of  $10^{13}$ Poise) at very low
temperature,      some     150    to    200K     below   the    actual
$T_{g}\thickapprox243K$: see figure \ref{fig:1}.  This points towards the
existence of a crossover temperature $T^{*}$ below which the dynamical
properties  of  a  liquid change  from  an  "ordinary", Arrhenius-like
temperature dependence to  an "anomalous", stronger than Arrhenius (or
super-Arrhenius) one. Such an anomalous  dependence, which is observed
in all  kinds of liquids  and polymers irrespective of  their detailed
molecular properties\cite{EAN96,DEBE01,TARKI01}, suggests a collective
phenomenon    involving  cooperative motion   of   a  large  number of
molecules.

\begin{figure}\begin{centering}
\resizebox{11cm}{!}{\includegraphics{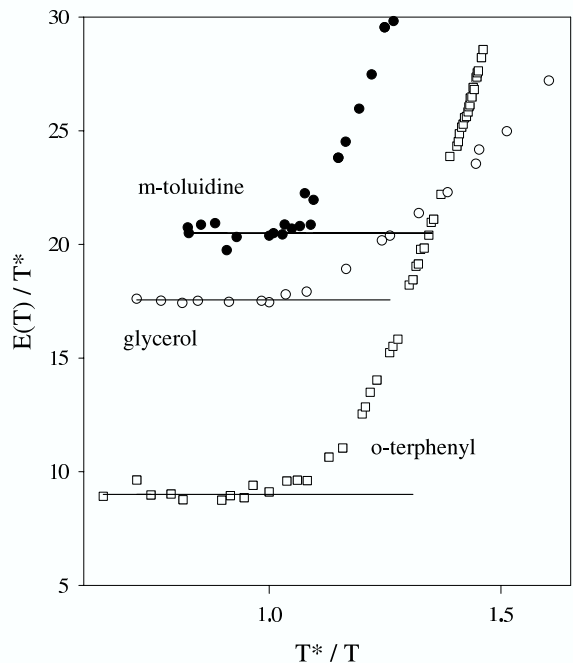}}
\caption{Crossover from Arrhenius to superArrhenius behavior for several
 glassforming   liquids:  effective  activation  (free) energy  $E(T)$
 versus   $1/T$.  Both $E(T)$ and  $T$   are divided by  the crossover
 temperature    $T^*$.     (Data  taken    from   references  cited in
 Refs.\cite{KTZK96} and \cite{A05}.)  } \label{fig:2}
\end{centering}\end{figure}

The notion of "fragility", introduced by Angell to classify glassforming
liquids and polymers\cite{CAA85}, characterizes the degree of super-Arrhenius
behavior, the more fragile the glassformer the greater the super-Arrhenius
character. The strong temperature dependence of the viscosity and
$\alpha$-relaxation time of glassforming liquids is often described by a
Vogel-Fulcher-Tammann formula, $\eta \approx \eta_{\infty}
\exp\left[DT_{0}/(T-T_{0})\right]$, $D$ being a measure of fragility. This
description implies the presence of a dynamical singularity at a temperature
$T_{0}$ lower than $T_{g}$. (However, $T_{0}$ is always found quite far below
$T_{g}$: at least 40 K for the fragile liquid ortho-terphenyl, see figure \ref{fig:1}.) An
alternative way of representing the data is to focus on the effective activation
(free) energy, $E(T)=k_{B}T \ln\left(\eta/\eta_{\infty}\right)$. In this
representation, the crossover behavior shows up quite clearly, as illustrated in
figure~\ref{fig:2} for several liquids.

Another characteristic of the approach  to the glass transition, which
becomes  more manifest  in the temperature  range  below the crossover
discussed  above, is  the   spatially  "heterogeneous" nature   of the
relaxation.  The most easily detected, yet not uniquely interpretable,
signature  of  such heterogeneity is   a  stretching of the relaxation
functions,    \textit{i. e.}, a    deviation  from  simple exponential
dependence at long times  (or  the equivalent broadening  in frequency
space for the dynamic susceptibilities); this  is often represented by
a stretched exponential in time, $\phi(t)\thicksim
\exp(-(t/\tau_{\alpha})^{\beta})$ where $\tau_{\alpha}$ is the
$\alpha$-relaxation  time.  More  direct  experimental   evidence  for  the
presence   of  supermolecular   dynamically  correlated  regions    in
supercooled liquids has been  provided in the  past ten years (see the
reviews \cite{Silles99,Edig00,R02}).    This, too, suggests collective
behavior; however, the    measured  (linear) size  of the   correlated
regions   never   exceeds  5     to   10   molecular   diameters    at
$T_{g}$\cite{A97}.

An additional puzzling feature of the glass transition problem is that
the spectacular change in the  dynamical properties does not come with
a              related         growth       in           thermodynamic
correlations\cite{EAN96,DEBE01,TARKI01}.   No  convincing evidence has
been  so far reported   for   a significant  increase  in   structural
correlations: the evolution of    the  static pair   structure  factor
appears weak  and featureless even in   very fragile glassformers. The
only noticeable   observation is the decrease  of  the excess entropy,
defined as the difference between  the entropy of a supercooled liquid
and that of   the   associated crystal, with  decreasing   temperature
\cite{K48}. This decrease appears correlated with  the slowing down of
the     relaxation,      being    larger     for    more    fragile
glassformers\cite{A97}

We can  then summarize  our view  of the  phenomenology of supercooled
liquids as follows:  glass formation seems to be  a {\it collective or
cooperative} (we use the two  terms interchangeably) phenomenon, which
would account  for  the  rather  universal  trends   observed  in many
different   systems, but   cooperativity   occurs on  a {\it  limited,
mesoscopic spatial scale}.

\begin{figure}\begin{centering}
\resizebox{11cm}{!}{\includegraphics{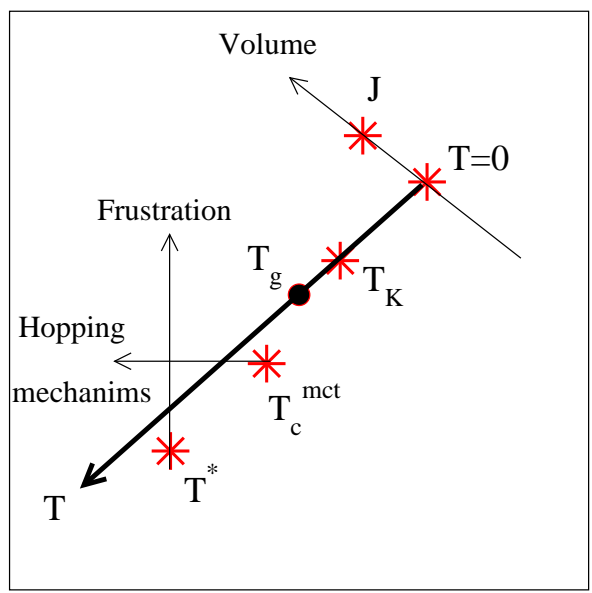}}
\caption{Schematic representation of  (critical) theories of the glass
transition in  a  virtual multidimensional diagram: the  physical axes
(temperature,    volume)   are  complemented by    abstract dimensions
(frustration,   hopping    mechanisms).     Avoided critical   points:
$T^*=T_c^{(0)}$,  critical point   in  zero frustration;  $T_c^{MCT}$,
dynamic singularity of  mode  coupling theory.   Unreacheable critical
points below $T_g$:   $T_K$,   entropy crisis or  random   first-order
critical point;    $T=0$, dynamic    critical point   for  kinetically
constrained  models; point   $J$,  jamming transition. (See  text  for
references.) }
\label{fig:0}
\end{centering}\end{figure}

As we  have  mentioned in the  Introduction, the  absence  of a unique
reading of  the salient aspects of   the phenomenology of glassforming
liquids goes with the absence of a widely accepted theory of the glass
transition. There is indeed a variety  of theoretical approaches which
often appear at  odds with one  another.   The one feature  that most
theories have in common, though, is the idea  that the slowing down of
relaxation and flow in  supercooled liquids (especially in the fragile
ones) is spectacular enough  and general enough to be  describable
by  a universal  theory  and that,  as   is quite natural  in physics,
universality is   associated with the  presence    of one  or  several
underlying   critical points: the  putative    critical points may  be
dynamic\cite{gotze91,BB04,GD03,WBG04}                               or
thermodynamic\cite{AG65,KTW89,XW00,MP97,KKNT95,SSH91,ON03}, but, as no
divergences or   singularities are actually   observed, those critical
points are postulated to be either unreachable
\cite{AG65,KTW89,XW00,MP97,GD03,WBG04,SSH91,ON03},
\textit{i.  e.}, occurring below  the experimental  $T_{g}$, or avoided
\cite{gotze91,KKNT95}, being  only present in  an ideal system "near"
to the real liquid (see figure ~\ref{fig:0}).

In  this article, we review   one such theoretical  approach, which is
based on the concept of {\it frustration}.  Frustration, meant here to
describe an  incompatibility between the  locally preferred order in a
liquid  and the global    requirements  for tiling   of space,  is  an
appealing candidate  for generating  collective (cooperative) behavior
on a  limited spatial scale and bring  in static  spatial correlations
only   at  a multi-particle level   that  cannot  be probed  by  usual
experimental techniques.

\section{Geometric frustration in simple atomic systems: the curved-space
approach}\label{sec:geom-frustr-simple}
\begin{figure}\begin{centering}
\resizebox{8cm}{!}{\includegraphics{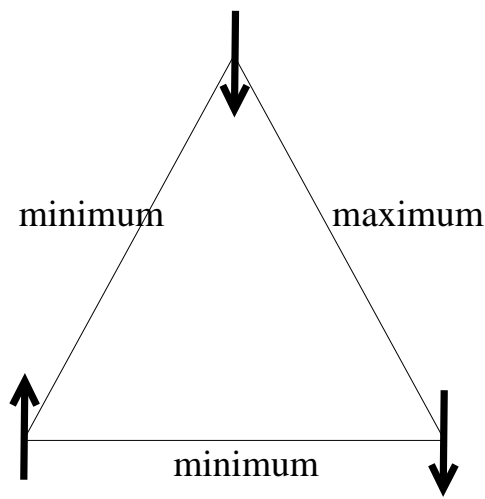}}
\caption{Elementary plaquette of a triangular lattice. When interactions 
between nearest neighbors Ising spins  are antiferromagnetic, one  can
never satisfy (i.e. anti-align spins) more than $2$  bonds at the same
time.} \label{fig:3}
\end{centering}\end{figure}

The  concept  of  frustration     was  introduced  by  Toulouse     in
1977\cite{G77} in the context of spin models to describe situations in
which    one cannot  minimize   the energy   of the  system  by merely
minimizing all local  interactions. A simple illustration  is provided
by an Ising spin model on a  triangular lattice with antiferromagnetic
interactions between nearest-neighbor spins: an elementary "plaquette"
of  the lattice  made by one    triangle of nearest-neighbor sites  is
frustrated because the  three antiferromagnetic interactions along the
bonds forming the plaquette can never be satisfied simultaneously (see
figure \ref{fig:3}).  Spin glasses are a well-known example of frustrated
systems; however, in this case  frustration is induced by the presence
of quenched  disorder, due for instance  to frozen-in impurities\cite{Y98}. This
inhomogeneous and externally imposed  frustration is not  relevant for
supercooled liquids in which glassiness and heterogeneous behavior are
self-generated. To   emphasize the   difference,  frustration in   the
context of liquids has been  given several qualifying adjectives, such
as  "uniform",  "geometric", "topological",  or "structural".  In what
follows however we shall  simply refer in  most cases to "frustration"
without adding any epithet.

\begin{figure}\begin{centering}
(a)\resizebox{7.5cm}{!}{\includegraphics{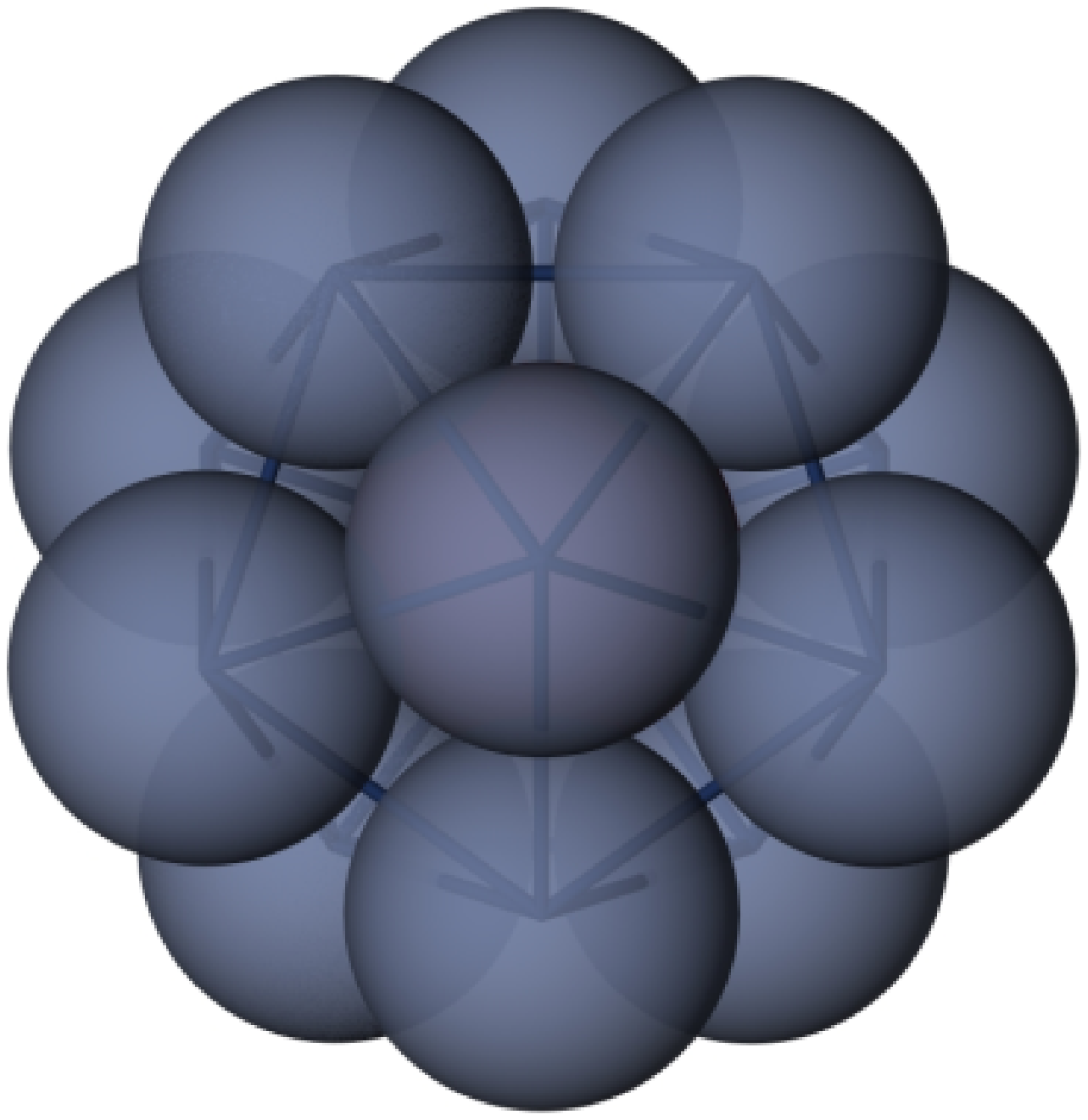}}
(b)\resizebox{7.5cm}{!}{\includegraphics{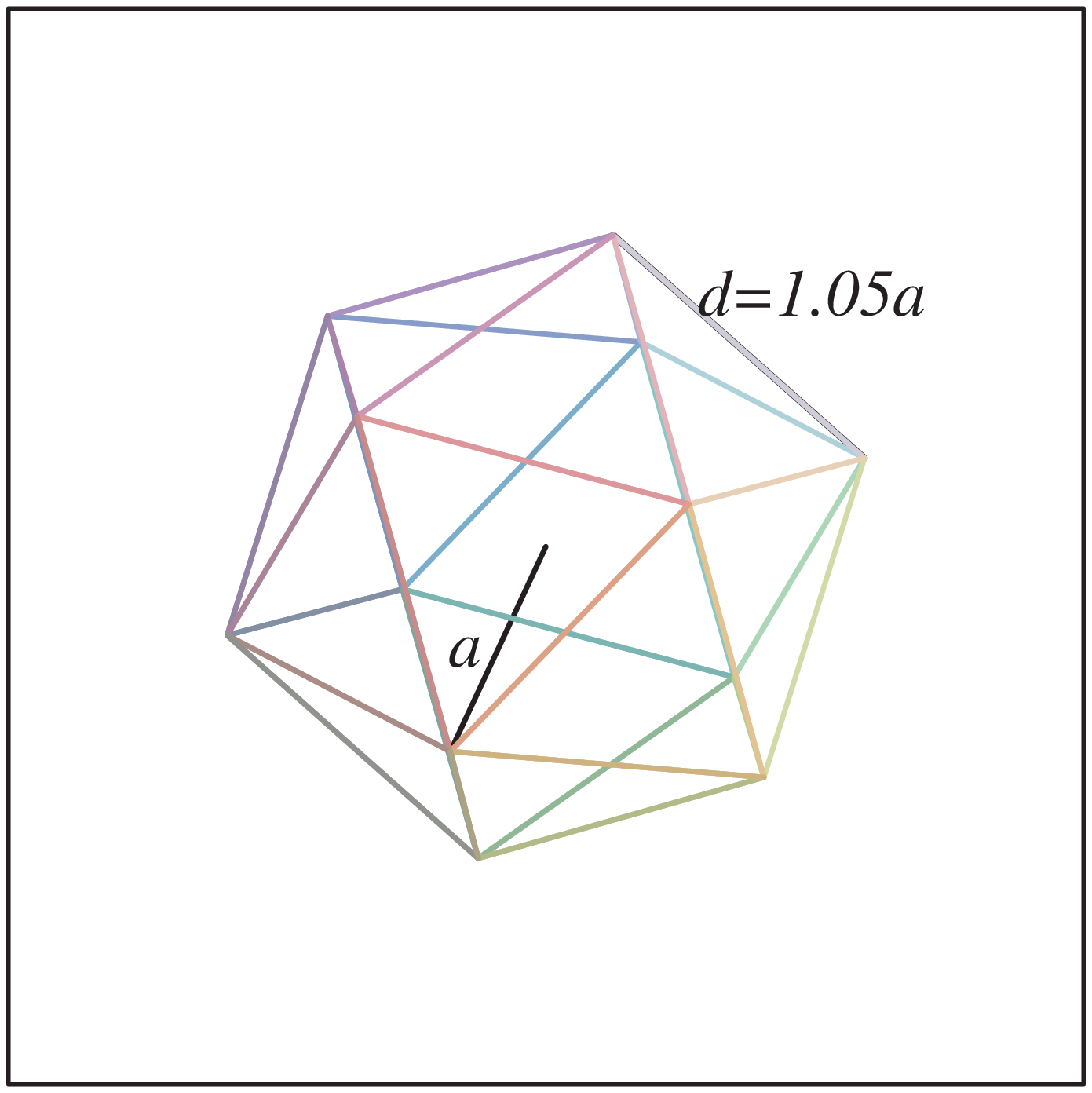}}\\
(c)\resizebox{7.5cm}{!}{\includegraphics{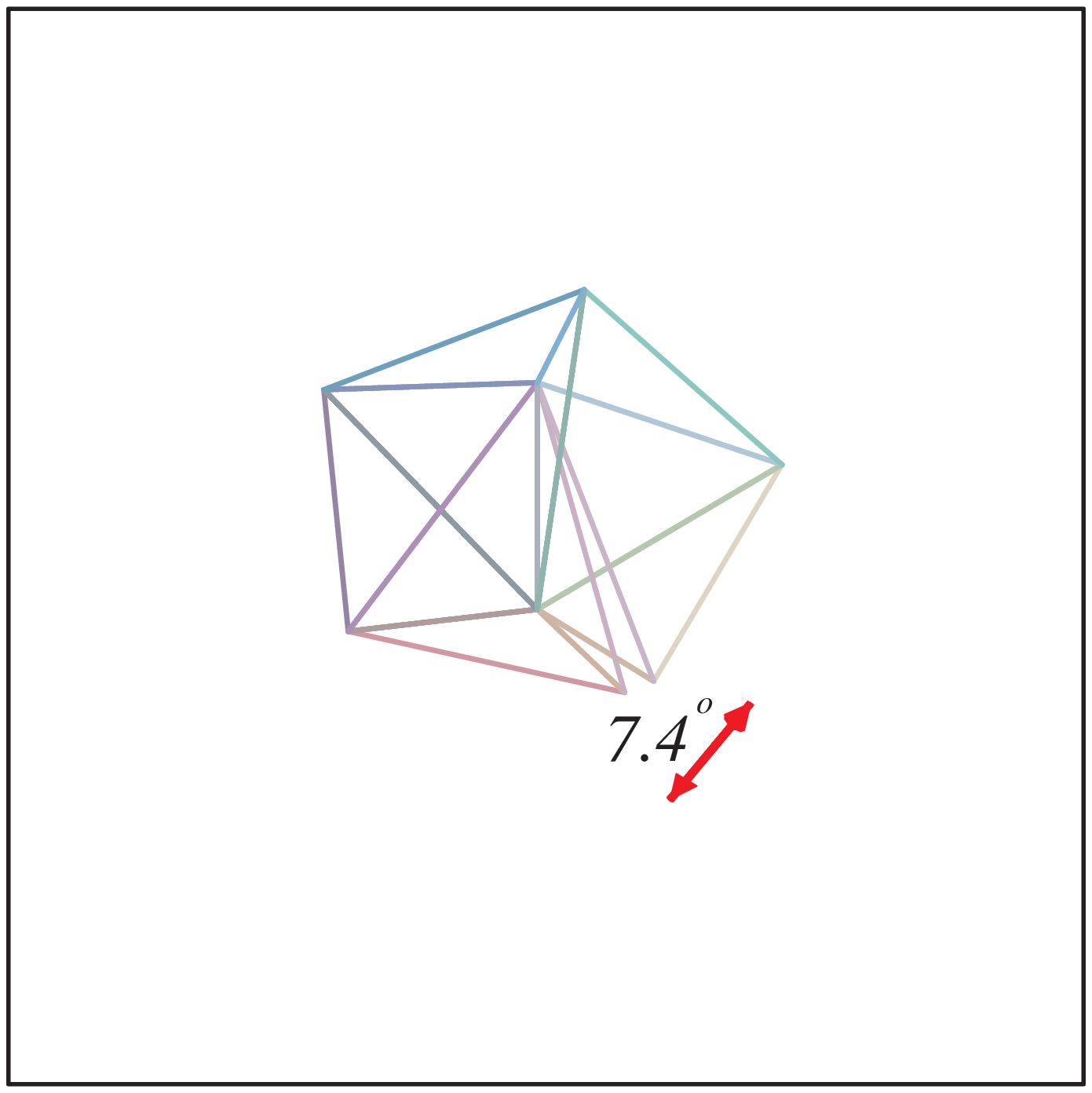}}

\caption{Manifestations of frustration for
tetrahedral/icosahedral order in   $3$-d atomic systems: (a) the  $5$-fold
rotational symmetry  of the icosahedron shown  here is  not compatible
with  translational periodicity. (b) The distance  d between the atoms
of the  first shell is slightly  larger than  the distance $a$ between
the central atom  and its  neighbors:  $d=1.05..a$. (c)  The  dihedral
angle of  the  tetrahedron, $\cos^{-1}(1/3)$, is not   commensurate with $2\pi  $ so  that $5$
tetrahedra packed along a bond leave a defect  angle of about $7.4^o$.} 
\label{fig:4}
\end{centering}\end{figure}

What  has often been taken as  the paradigm of (geometric) frustration
in  amorphous systems is   the case of    icosahedral order in  simple
one-component liquids in  which the atoms interact through spherically
symmetric pair potentials. The ground state of four atoms is a perfect
tetrahedron, with the atoms sitting at  the vertices, and  twenty such
tetrahedra  can  be  combined  to form a  regular  13-atom icosahedron
(figure \ref{fig:4}).   Frank\cite{F52}  was  the  first  to   stress the
importance of  local icosahedral  order  in simple atomic  liquids and
liquid  metals.  He showed  that the  most  stable  cluster made of  a
central atom and a shell of 12 neighbors is indeed an icosahedron, and
not an arrangement associated with the actual crystalline phases, bcc,
fcc,  or hcp. In  3-dimensional Euclidean space, however, this locally
preferred  structure  (tetrahedral   or icosahedral) cannot  propagate
freely to   tile the  whole space:  this   is what  is  meant  by {\it
geometric frustration}.  The  global  ground state of   the  system is
instead an fcc or hcp crystal.

The concept of frustration is more easily  grasped by constrasting the
case  of  spherical   particles   in     3  dimensions   with    other
situations\cite{N02,SM99}.   Systems    of    spherical particles   in
2-dimensional Euclidean space, \textit{i.  e.}  disks on a plane,  are
not subject  to  frustration:  the  locally preferred   structure is a
regular hexagon, with  one atom at the  center and 6 neighbors  at the
vertices, and  this structure can be periodically  repeated to  form a
triangular lattice.  Similarly,   aligned hard cubes in  3-dimensional
Euclidean space form simple  cubic arrangements that are  both locally
and   globally   preferred,    therefore   excluding  any  frustration
effect\cite{J98}.

As illustrated in figure \ref{fig:4}, frustration  can be envisaged from various   different  angles: as a   global  constraint,  e.   g., the
incompatibility  between  the   5-fold  rotational   symmetry  of  the
icosahedron and translational periodicity (figure \ref{fig:4}a), or, more
in line with  the usual definition  of frustration (see above), as the
impossibility  to let all  the atoms  simultaneously sit in  positions
corresponding to the minima   of the pair potentials  between  nearest
neighbors (figure \ref{fig:4}b).

An  important observation is that  frustration  can be  turned off  by
"curving" space\cite{KS79,SM99}.   One can play  with the  metrics and
topology  of the  underlying space in  order to  let   the local order
propagate freely and generate a regular tiling or "tessellation". This
can be understood  for instance by considering figures \ref{fig:4}b and
c: if  space  can be curved so   that  the defect  angle  made by  $5$
tetrahedra packed along a bond as well as the mismatch between the two
interatomic  distances $d$     and  $a$  in  an  icosahedron   vanish,
propagation    of   tetrahedral/icosahedral order    becomes possible.
Indeed,  a perfect  icosahedral  phase  can be  formed  on the surface
$S^{3}$  of a  4-dimensional  hypersphere with   a radius tuned  to be
$5/\pi=1.5915...$ times the preferred  interatomic geodesic distance (in
curved  space).     This ideal  icosahedral  structure   is called the
polytope $\left\lbrace3,3,5\right\rbrace$\cite{SM99,C91}.

The core of the curved-space approach of amorphous (liquid and glassy)
phases,    developed  on  one    hand     by  Sadoc,   Kleman,     and
Mosseri\cite{KS79,SM99,MS90,Sa83,SM84}  and   on  the other   by Nelson and
coworkers\cite{N83b,N83,N02,NW84,SN84,SN85}                        and
Sethna\cite{Se83,Se85}, consists  in  using this ideal  structure as a
reference state (for reviews see Refs.\cite{N02,SM99,VD85,VD86}.  Going
back from the reference state  to the actual configurations in  "flat"
(Euclidean)  space   necessarily forces  in  topological defects  that
perturb the ideal order. All sorts of defects can be generated, but it
was argued  that the most relevant for  the physics  of atomic glasses
are "disclinations"\cite{Sa83,SM99,N83b,N83}: disclinations are
associated with the  breaking of  rotational, here bond-orientational,
symmetry (their    counterpart  for translational   symmetry   are the
"dislocations"),   and  they    are        line  defects   in        3
dimensions\cite{K89}. Such topological defects  are expected to form a
disordered network, the    actual amorphous phases being composed   of
locally ideally ordered regions interrupted by defects
\cite{Note5}.  Interestingly,  the network of  defects (disclinations)
may itself become   ordered at  low  temperature, thereby   leading to
"defect-ordered phases"  whose best examples in  the case  of metallic
systems are the Frank-Kasper phases\cite{SM99,FK58,FK59}.

Whereas Sadoc and Mosseri focused on the most efficient way to produce
dense      amorphous      packings    by    decurving     the polytope
$\left\lbrace3,3,5\right\rbrace$,             Nelson               and
coworkers\cite{NW84,SN84,SN85,N02} developed a  statistical mechanical
approach of frustrated icosahedral  order. They built  a set  of local
order  parameters $\mathbf{Q_{\ell}(\mathbf{x})}$ by  projecting a  local
particle configuration   centered on  position $\mathbf{x}$   onto the
surface of a tangent 4-dimensional hypersphere with appropriate radius
to accommodate the polytope $\left\lbrace3,3,5\right\rbrace$ and, much
in the spirit of the  Ginzburg-Landau theory of freezing, they derived
a free-energy functional of this local order parameter\cite{NW84},

\begin{equation}\label{eq:1} 
F[\mathbf{Q_{\ell}}]=\int d^{3}x \lbrace \frac{1}{2}Z_{\ell}
\mid D_{\mu}\mathbf{Q_{\ell}}(\mathbf{x})\mid^{2} + \frac{1}{2}
\tau_{\ell} \mid\mathbf{Q_{\ell}}(\mathbf{x})\mid^{2} +
O(\mathbf{Q_{\ell}}^{3}) \rbrace,
\end{equation}

where  the  index $\mu$ is  associated with  the spatial  coordinates in
ordinary  $3$-D  Euclidean space  (e.     g.,  $\mu=x,y,z$ for   cartesian
coordinates); the  local order  parameter $\mathbf{Q_{\ell}}(\mathbf{x})$
transforms according to the representation  of dimension $(\ell + 1)^{2}$
of $SO(4)$, $Z_{\ell}$ and $\tau_{\ell}$  are phenomenological parameters,  and
$O(\mathbf{Q_{\ell}}^{3})$  denotes  cubic   and higher-order  invariants
built  with $\mathbf{Q_{\ell}}$.  The  most  relevant value   of  $\ell$ for
describing  icosahedral order is  12, but $\ell=20, 24,  30, 32, ...$ are
also allowed  by  symmetry  ($\ell=0$ corresponds   to the  mean particle
density).

Frustration is introduced via   the so-called  "covariant  derivative"
entering   in the  gradient  term.   Following a  suggestion  made  by
Sethna\cite{Se83,Se85},  it is defined by  requiring that the gradient
term be  minimum when  two nearby  local  particle  configurations are
related by "rolling without slipping" the tangent hypersphere carrying
the   reference  icosahedral  template along   the   path joining  the
configurations.  This leads to\cite{NW84}

\begin{equation}\label{eq:2} 
D_{\mu}\mathbf{Q_{\ell}}(\mathbf{x})=\partial_{\mu}\mathbf{Q_{\ell}}(\mathbf{x})
 -  i\kappa
\mathbf{\hat{L}^{(\ell)}}_{0\mu} \mathbf{Q_{\ell}}(\mathbf{x}) ,
\end{equation}

where the $\mathbf{\hat{L}}^{(\ell)}_{0\mu}$'s denote the generators of the
$SO(4)$  rotations in the $(0,\mu)$  plane ($0$ denotes the direction in
the  $4$-dimensional space  embedding $S^3$ which  is perpendicular to
the    tangent Euclidean space  described   by the cartesian coordinates
$\mu=x,y,z$) and  $\kappa$ is the inverse  radius of  the hypersphere $S^{3}$
compatible  with the polytope $\left\lbrace3,3,5\right\rbrace$.   Here
and in the  following, the Einstein  convention is used, in  which one
sums over indices whenever they are repeated. The covariant derivative
is  frustrated because  it  cannot be  minimized (\textit{i. e.},  set
equal to zero) everywhere.

Despite its many appealing features, this curved-space approach of frustrated
icosahedral order has led in practice to meager results concerning the glass
transition phenomenon itself. Several reasons may explain this shortcoming.
First, the models, such as the $SO(4)$ uniformly frustrated theory described
above, are rather complex and have never been fully analyzed: for instance, the
topological defects have been characterized\cite{N83,N83b,NW84} and the static structure
factor of metallic glasses has been reproduced\cite{SN84,SN85}, but no attempts
have been made to study the dynamics.

Secondly, difficulties may arise from the
fact that local icosahedral order in simple atomic systems appears strongly
frustrated. The typical distance between defects, expected to be of the order
or less than the radius of the reference hypersphere $\kappa^{-1}$, is small:
one or two particle diameters at most, which precludes a separation between a
mesoscopic, coarse-grained description of collective effects and a detailed
account of the microscopic properties. (This may also explain the controversy
about the extent of icosahedral order and associated correlations in amorphous
phases composed of spherical particles.) It is worth noting that simple atomic
systems are not good glassformers and appear quite "non-fragile" with an
essentially Arrhenius temperature dependence of the viscosity and
$\alpha$-relaxation time\cite{FFMMNRSST03}; this is also true for the Lennard-Jones
models studied in computer simulations which, in spite of the weakness of the
interatomic bonding, are much less fragile than typical molecular glassforming
liquids\cite{TKV00}.

Finally, if glass formation is indeed a "universal"
phenomenon whose most salient properties are largely independent of molecular
details (in a sense discussed in the preceding section), icosahedral order is
likely to be too material-specific. Possible generalizations to describe
frustration in liquids are reviewed in the next section.

\section{Statistical mechanics of frustration in liquids}\label{sec:stat-mech-frustr}

Application of the concept of frustration to liquids boils down to three 
propositions  which  are plausible and, as discussed above, reasonably  well
established in the case  of systems of spherically  symmetric particles, but
still remain at the level of postulates in the case of molecular liquids,
mixtures, and polymers. Those propositions can be summarized as follows:

1) \textit{A liquid is characterized  by a locally preferred structure
(LPS) which is different than that of the crystalline phases.} This LPS
is an arrangement  of molecules that minimizes  some local free energy
(an operational definition  of  such a local free  energy  is given in
Ref.~\cite{MT03}),      much   in   the  same    way     as    the local
tetrahedral/icosahedral      order    in  liquids     of     spherical
particles.  Description   of such a  LPS   requires  the  knowledge of
multi-particle densities,  beyond the  usual one-body density,  and of
multi-particle correlations, beyond the  usual pair correlations.  For
instance, even  in simple    atomic  liquids, information  on    local
bond-orientational order involves at least $2$-body densities and $4$-body
correlations\cite{SNR81,SNR83,S92}. For this  reason,  the signature of  a
LPS is hard to  detect in common diffraction  experiments in which one
has only access to the (pair) structure factor of a liquid or a glass.

The LPS  may  change at high  enough  pressure: this is the   case for
instance of water and    tetrahedrally bonded liquids for  which   the
low-pressure LPS that tends to minimize the local energy by satisfying
the maximum number  of directional (hydrogen-  or covalent) bonds goes
over to a different  structure at high  pressure where packing effects
dominate over bonding effects \cite{YABW96,MS98,TAGVK03}. 
However, at a given pressure or density, the LPS does not change with 
temperature; it just becomes more and more favored as temperature decreases.

2) \textit{The  LPS characteristic of a given  liquid cannot  tile the
whole space}. This   incompatibility  between local order  and  global
space  filling is   precisely  the    content  of  the  concept     of
\textit{frustration}. Were it   not  for frustration,   the  LPS could
propagate through the  whole space and form  an  "ideal" ordered phase
that  would supersede the actual  crystalline phases. Instead of this,
the actual crystal is formed via  a strong first-order transition that
requires a rearrangement of the local structures,  a cost that is more
than compensated  by the free-energy gain due  to the global tiling of
space.  So, as   suggested  by    Frank\cite{F52},  the presence    of
frustration   is the   possible  physical  ingredient   allowing
supercooling of a liquid at temperatures below the melting point.

3) \textit{It is possible to construct an abstract reference system in
which the  effect of  frustration is   turned off}. At  a  microscopic
level,  frustration  implies that  the  atoms or the  molecules cannot
simultaneously  sit  in the minima of   all pairwise interactions with
their nearest  neighbors. By allowing one to  modify the  topology and
the  metrics  of  the  underlying     space,  one  can   get rid    of
frustration. This is  the  essence  of  the curved-space  approach  of
icosahedral order (see above). In an abstract parameter space, one can
now think  of frustration as a  tunable  control variable.  However, a
given  liquid  is    of  course characterized  by  a    given level of
frustration, a given LPS, and a given reference space.

Note also  that in the  absence  of  frustration, "ideal" ordering  is
likely to proceed via  a continuous or weakly  first-order transition,
quite differently  than the standard  melting  in $3-D$ flat  space. For
instance,  the melting transition of parallel  cubes, a system with no
frustration  between  local order  and  global tiling in  ordinary $3-D$
space, has been shown   to be continuous\cite{J98}.  Similarly, Nelson
and   his  coworkers\cite{NW84,SN84,SN85,N02}  have       argued  that
bond-orientational ordering  is likely to   occur via a continuous  or
weakly first-order transition in 3 dimensions, even in the presence of
cubic     invariants   in     the     Landau    description   of   the
system\cite{NT81,SNR83}.  (Icosahedral ordering in the polytope $\left\lbrace  3,3,5\right\rbrace  $ on $S^{3}$\cite{S84} does not occur by a true phase transition at all since the underlying space is the surface of a 4-D sphere of finite radius.)

The   next step is to  build   a statistical mechanical description of
glassforming liquids  based  on the above   propositions. We will take
here the point of  view that if indeed,   as argued in discussing  the
phenomenology, the viscous slowing down of supercooled liquids and the
resulting glass transition are   universal,  progress can be made   by
relying on a coarse-grained, mesoscopic approach that incorporates the
effect of frustration while,  at least in a  first step, overlooking a
detailed description  of the LPS,  of the reference  space, and of the
precise mechanism   by which frustration operates\cite{Note8}.

A \textit{minimal   theoretical model} of  frustration in  glassforming liquids
should satisfy,  on top of the  overall consistency  with the physical
picture developed above,   some prerequisite properties,  such  as (i)
frustration should   be  a  tunable parameter,   which allows  one  to
investigate   the   potential  connection  between   frustration   and
"fragility" in glassformers,  and (ii) frustration should be  uniform,
\textit{i.e.}, the same at every point of space and, consequently, not
due to the presence of quenched disorder.  We review below some of the
paths that have been, and still are, followed. (Note that in all these
descriptions,  the actual crystal  is excluded  from  the picture,  in
order to focus on the liquid and supercooled liquid phases.)

\subsection{\textbf{Coupling to a non-Abelian gauge background}}

The statistical mechanical treatment of  icosahedral order put forward
by  Nelson and his  coworkers (see above) can be  extended by ignoring
the precise
reference of the local order  parameter (and of the  LPS) to the ideal
tiling   of the    polytope $\lbrace3,3,5\rbrace$ and    that   of the
frustration strength  to  the radius of  curvature  of the hypersphere
$S^{3}$ compatible with this ideal tiling.

In a continuum, field-theoretical description, the generic form of the
free-energy functional reads\cite{N04}

\begin{equation}\label{eq:3} 
F[\mathbf{Q}]=\int    d^{3}x        \lbrace            \frac{1}{2}    Z \mid
D_{\mu}\mathbf{Q}(\mathbf{x})\mid^{2}            +            \frac{1}{2} \tau
\mid\mathbf{Q}(\mathbf{x})\mid^{2} + O(\mathbf{Q}^{3}) \rbrace,
\end{equation}

where the index   $\mu$ is associated  with  the spatial coordinates  in
ordinary $3-D$    Euclidean  space;    the  local order   parameter $\mathbf{Q}(\mathbf{x})$
transforms according to the representation of dimension $n$ of a given
non-Abelian   continuous Lie group, $Z$ and $\tau$ are
phenomenological parameters, and $O(\mathbf{Q}^{3})$ denotes cubic and
higher-order invariants built with $\mathbf{Q}$.

The "covariant derivative" entering in the gradient term is defined as

\begin{equation}\label{eq:4} 
D_{\mu}\mathbf{Q}(\mathbf{x})=\partial_{\mu}\mathbf{Q}(\mathbf{x}) - ig
A_{\mu}^{\alpha}
\mathbf{\hat{L}_{\alpha}} \mathbf{Q}(\mathbf{x}) ,
\end{equation}

where  the  $\mathbf{\hat{L}_{\alpha}}$'s   denote the  generators   of the
representation of  dimension   $n$  of   the  considered   non-Abelian
group. The $A_{\mu}^{\alpha}$'s  are coefficients depending on both "spatial"
indices $\mu$ and  ``internal'' indices  $\alpha$;  they are independent   of
position and play  the  role of  a  frozen, uniform  gauge  field (see
below); the  coupling strength $g$  characterizes the magnitude of the
frustration.  (We recall  that the  Einstein  convention is  used  for
summing indices.)

The   thermodynamic and structural  properties of  the system can been
obtained from the partition function

\begin{equation}\label{eq:5} 
{\cal Z} = \int  \mathcal{D}\mathbf{Q} exp\left[-\beta F[\mathbf{Q}]\right] ,
\end{equation}

where $\beta =1/(k_BT)$, whereas studying  the relaxation properties requires  the introduction
of some local dynamical rules that reflect, at a coarse-grained level,
the underlying motion of the atoms or molecules.

For   example, in  the theory  developed   by Nelson and summarized in
section~\ref{sec:geom-frustr-simple}, the non-Abelian  group        is
$SO(4)$, the  main  local   order parameter  transforms   according to  the
representation of $SO(4)$  associated  with angular   momentum $l=12$,
\textit{i. e.},  of dimension $n=(l+1)^{2}=169$,   $\alpha$ denotes the $6$
rotation planes in  4-D   space, $(0,x), (0,y), (0,z),  (x,y),  (x,z),
(y,z)$; $A_{\mu}^{\alpha}$ is equal to one when $\alpha$ corresponds to the planes
$(0,\mu)$  with $\mu=x,y,z$   and zero  otherwise, and   $g$  is  equal to
$\kappa_{*}$,   the inverse  of   the  radius   of the "ideal"    reference
hypersphere.   A  related,   but somewhat  simpler    approach  due to
Sethna\cite{Se83,Se85} corresponds to the $l=1$ (\textit{i. e.},
$n=4$) representation of $SO(4)$ (see  also Ref.~\cite{N04}).  Lattice
versions of the above free-energy functional have also been proposed
\cite{CafLB86,N04}, which  could  make possible  computer simulation
studies of such theories.

In  the present description,  frustration is embodied  in the specific
form of the  covariant derivative. When  the coupling strength $g$  is
set  to  zero, the reference, unfrustrated   model is chosen such that
there   is  a   continuous or   weakly  first-order   transition to  a
symmetry-broken  phase in which  the  order parameter takes a non-zero
value (this  phase describes the  ideal order  formed by extending the
LPS to the whole space). When  $g$ is different from zero, frustration
arises from the fact that the covariant derivative cannot be made zero
everywhere. For instance, if  one builds configurations by forcing the
covariant  derivative to vanish along   straight lines starting in all
directions  from a  given  position $\mathbf{x}$, one   finds that the
covariant  derivative differs  from   zero along  any closed   circuit
encircling $\mathbf{x}$;  this  leads  to  a large  free-energy   cost
associated with a non-zero  gradient term. (Actually, as  discussed in
more  detail below, this  cost grows super-extensively with the linear
size of the region around  the arbitrary chosen point $\mathbf{x}$, up
to some intrinsic frustration length set up by the value of $g$.)

In the  framework   of differential  geometry\cite{LL60,SM99,K89,VD86},  making   the covariant
derivative  vanish along a  curve amounts to parallel transporting the
local  order parameter (which  is  a vector or  a  tensor) along  this
curve. In the curved-space approach of Sethna and Nelson, this is done
by "rolling"  a  tangent 4-D sphere $S^3$   with  the ideal icosahedral
template  along the chosen  curve in Euclidean space. Frustration then
means  that one cannot  extend such a parallel  transport to the whole
space.  The reason  is that the   covariant derivative has a  non-zero
curvature; this curvature is directly related to  the magnitude of the
difference between  the initial and  final values  of the local  order
parameter  after   parallel transport  along    a closed circuit.  The
presence  of a  non-zero  curvature   (not to be  confused  with   the
curvature of the physical Euclidean space  which is of course zero: in
the Sethna-Nelson picture,  the non-zero  curvature of the   covariant
derivative  is precisely equal  to  that of the  reference  4-D sphere)
comes   from       the     non-commutativity   of     the     generators
$\mathbf{\hat{L}_{\alpha}}$, \textit{i.  e.}, from the non-Abelian property
of the gauge background.

An  important consequence of frustration  is the necessary presence of
defects:   the   ideal order  that  forms the     ground  state of the
unfrustrated system ($g=0$) can no longer propagate freely through the
whole  space, and    low-temperature configurations must    contain an
irreducible density   of defects (e.  g., disclination  and dislocation
lines,  grain boundaries, etc...). Such   "topological" defects can be
studied by means  of the homotopy  theory\cite{M79}, as done by  Nelson
and Widom\cite{NW84} for their $SO(4)$ theory of local icosahedral order.

For completeness,  it is worth interpreting  the above class of models
in the language of gauge  field theory\cite{K89,VD86}. The local order
parameter $\mathbf{Q}(\mathbf{x})$ represents the matter field that is
"minimally   coupled" to the  non-Abelian  gauge field. In the present
case, the  gauge field is  frozen  and uniform  (hence the term "gauge
background" used to describe such  a theory). The associated  physical
field,  defined as   the  so-called covariant curl  $F_{\mu\nu}=-i  \left[
D_{\mu}, D_{\nu}  \right]$, is then   given by $F_{\mu\nu}=i  g^{2}  A_{\mu}^{\alpha}
A_{\nu}^{\beta} [ \hat{L}_{\alpha}, \hat{L}_{\beta} ]$. It is non-zero because of the
non-commutativity property of  the $\hat{L}_{\alpha}$'s associated with the
non-Abelian nature  of the "gauge  group" under consideration. Contact
can be  made with the   formalism of differential geometry and  tensor
calculus by  noticing  that  $F_{\mu\nu}$ identifies  with  the  curvature
tensor\cite{K89}. Note that the present theory is  only globally invariant under
transformations of this group. If necessary,  a full blown gauge field
theory in which local invariance   is now enforced via a  fluctuating,
space-dependent gauge field  could  be  developed; attempts in    this
direction have been made \cite{RD82,R83,K93,K02}.

\subsection{\textbf{Uniformly frustrated spin models}}

In developing their approach based on geometric frustration for simple
liquids and glasses, Nelson and his coworkers\cite{N83,N02,SN85} proposed an
analogy with
the effect of an applied,  uniform magnetic  field in extreme  type-II
superconductors\cite{T80}.  In  such systems,  their is  always at least partial
penetration of the magnetic field, no matter how small the field: this
latter  induces an irreducible   density of   defects (flux  lines  or
vortices) of the  same "sign". The ground  state is then an  Abrikosov
lattice in which those  vortices form a  periodic array, in  a similar
way as Frank-Kasper phases in which disclination lines form a periodic
network are the expected ground states of simple metallic systems. The
curvature mismatch  that  generates  frustration in the   curved-space
approach is replaced here by the applied magnetic field.

The analogy can be  made somewhat more  precise in two dimensions.  As
discussed   in    section~\ref{sec:geom-frustr-simple},   there  is no
frustration in that case for liquids formed by one-component particles
interacting via  spherically symmetric  pair potentials: the hexagonal
LPS can propagate through the whole space to form a triangular lattice
and   the freezing transition  is  either weakly first-order or splits
into two continuous transitions  separated by an  intermediate hexatic
phase characterized by bond-orientational  order but no  translational
order.  Geometric frustration can  now be introduced by curving space,
\textit{i. e.}, by  placing the liquid  on a hyperbolic plane $H^2$, a
surface  of   constant negative  curvature   $-\kappa^{2}$\cite{N02,RN83,N83}.  Let  focus  on
bond-orientational order. Frustration then comes from the fact that if
one   measures  the bond  angle  formed    by the vector   joining two
neighboring  atoms with respect  to  a reference axis, this  reference
axis   changes  when   it  is  parallel   transported  on  the  curved
surface. (On a    curved  manifold, there is   no   global  notion  of
parallelism,  so that   comparison of  reference   frames at different
points in  space requires  a  rule for  parallel transport  of vectors
along curves.)   A non-zero curvature implies  that the reference axis
does not  go back to  its  original value  when transported along  any
closed circuit.  In  the  long wavelength, continuum   approach  (also
called  "hydrodynamic"  in     this  context),  the    Ginzburg-Landau
free-energy functional describing  local hexatic order on a hyperbolic
plane can be written as a modified gradient term\cite{N83},

\begin{equation}\label{eq:6}
F[\mathbf{n}]=\frac{1}{2}    Z_{H} \int   d^{2}x   \sqrt{g(\mathbf{x})} \mid
D_{\mu}n^{\nu}(\mathbf{x})\mid^{2} ,
\end{equation}

where $\mathbf{n}(\mathbf{x})$   is  a unit  vector tangent to  a
"bond"  centered   at   point  $\mathbf{x}$,  $g(\mathbf{x})$   is the
determinant   of  the   metric   tensor $g_{\mu\nu}(\mathbf{x})$ that   is
appropriate for the coordinate system chosen for the hyperbolic plane,
$Z_{H}$  is  a phenomenological  hexatic   stiffness constant, and the
covariant derivative is given by

\begin{equation}\label{eq:7}
D_{\mu}n^{\nu}(\mathbf{x})=\partial_{\mu}n^{\nu}(\mathbf{x})                    
   +
\Gamma_{\lambda\mu}^{\nu}(\mathbf{x})n^{\lambda}(\mathbf{x}) ,
\end{equation}

where the $\Gamma_{\lambda\mu}^{\nu}$'s are the  so-called connection components
that
define   parallel  transport  on $H^2$;    here   they are simply   the
Christoffel symbols obtained from derivatives of the metric tensor and
associated   with  the Levi-Civita  connection.  In  a local cartesian
coordinate  system   centered  at  an  (arbitrary)   origin,  one  has
$g_{\mu\nu}(\mathbf{x}) = \delta_{\mu\nu} - \kappa^{2} x_{\mu} x^{\nu} +
O(\kappa^{4} x^{4})$ and
$\Gamma_{\lambda\mu}^{\nu}(\mathbf{x}) = \Gamma_{\mu\lambda}^{\nu}(\mathbf{x}) =
-\kappa^{2} \delta_{\lambda\mu} x^{\nu}
+ O(\kappa^{4} x^{3})$ for   distances much less  than the  frustration, or
curvature, scale, $\kappa^{-1}$.

Consider now the Ginzburg-Landau free-energy functional for an extreme
type-II superconducting film in a perpendicular magnetic field $B$\cite{T80}:

\begin{equation}\label{eq:8}
F[\psi]=\int d^{2}x \lbrace \frac{\hbar^{2}}{2m} \mid  (\partial_{\mu} -
\frac{2 e  i}{\hbar c}
A_{\mu}(\mathbf{x}))\psi(\mathbf{x})\mid^{2} + 
\frac{1}{2} \tau \mid \psi(\mathbf{x})\mid^{2}+ O(\psi^{4}) \rbrace  ,
\end{equation}
 
where  $\psi(\mathbf{x})$ is the  complex superconducting order parameter
and $\mathbf{A}(\mathbf{x})$ is   the  vector potential  whose   curl,
$\mathbf{\partial} \land \mathbf{A} =  \mathbf{B}$,  is a fixed  constant  (where
$\mathbf{B} = B \hat{\bf z}$ is perpendicular to the plane $(x,y)$).

Rewriting $\psi(\mathbf{x})$ in  terms of two real fields, $\psi(\mathbf{x})
= \psi^{1}(\mathbf{x})  +  i \psi^{2}(\mathbf{x})$, and dropping terms other
than the (modified) gradient one, Eq.(\ref{eq:8}) can be recast as\cite{N83}

\begin{equation}\label{eq:9}
F[\mathbf{\psi}]=\frac{\hbar^{2}}{2m} \int  d^{2}x \mid 
(\partial_{\mu}\psi^{\nu}(\mathbf{x})  +
\Gamma_{\lambda\mu}^{\nu}(\mathbf{x})\psi^{\lambda}(\mathbf{x}))\mid^{2} ,
\end{equation}

with the "connection"  now defined as $\Gamma_{\lambda\mu}^{\nu}(\mathbf{x})=(2 e i/\hbar
c)  A_{\mu}(\mathbf{x}) \epsilon_{\lambda \nu}$,   $\epsilon_{ \lambda \nu}$    being      the   usual
antisymmetric   tensor with $\epsilon_{12}=-\epsilon_{21}=1$ and $A_{\mu}(\mathbf{x})$
being equal to $A_{\mu}(\mathbf{x})=-(1/2)B\epsilon_{\mu\sigma }  x^{\sigma }$, due to  the
uniform property of  the  magnetic field $B$  (see above).   Comparing
Eqs.(\ref{eq:6},\ref{eq:7})  with    Eq.(\ref{eq:9}) and    using  the
expansions of  $g_{\mu\nu}(\mathbf{x})$  and $\Gamma_{\lambda\mu}^{\nu}(\mathbf{x})$  for
distances less than the frustration scale, one sees a great similarity
between the two  systems, with $B$  playing  the role of $\kappa^{2}$;  the
mapping however  is  not exact  because the  tensorial  content of the
various objects is not  quite the same (in  particular the gauge group
$U(1)$ associated with superconductors   is Abelian, which is  not the
case for that  associated with  the connection  on $H^2$)  and because
curvature introduces an intrinsic   scale  $\kappa^{-1}$ beyond which   the
correspondence breaks down.

The analogy between the  two kinds  of  systems is also striking  when
looking at a dual picture in terms of defects. The relevant defects in
the case of  hexatic ordering are disclinations,  which are points  in
two dimensions.  When frustration is weak,   \textit{i. e.}, for small
curvature,   and at low  temperature,   it has  been shown that  local
hexatic order on  a curved surface  (here, the hyperbolic plane  $H^2$)
can be  described     in the continuum,  "hydrodynamic"     limit by a
free-energy    functional   for  the   defect (disclination)   density
$s(\mathbf{x})$,\cite{PL96,BNT00}

\begin{equation}\label{eq:10}
\fl
F[s]= N  E_{core} + \frac{1}{2}  Z \int d^{2}x  \sqrt{g(\mathbf{x})} \int
d^{2}y                                \sqrt{g(\mathbf{y})}
(s(\mathbf{x})+\kappa^{2})G(\mathbf{x},\mathbf{y})(s(\mathbf{y})+\kappa^{2}) ,
\end{equation}

when the  total number of  defects is equal to  $N$, $E_{core}$ is the
energy penalty associated with creating one defect, and the density of
defects                        is              defined              as
$s(\mathbf{x})=(\pi/3)\sum_{i=1,..,N}q_{i}\delta(\mathbf{x}-\mathbf{x_{i}})$,
with   $q_{i}=\mp 1,\mp 2,...$  the   topological   charge of  the  defects;
$G(\mathbf{x},\mathbf{y})$ is the  inverse Laplacian on the hyperbolic
plane,  which goes as the  coulombic  interaction $\ln(r/a_{0})$, with
$r$ the distance between two defects and  $a_{0}$ the core size of the
defects, up to a screening  length played by the intrinsic frustration
or  curvature scale  $\kappa^{-1}$.  The  above  expression is  obtained by
requiring    "charge  neutrality",  which    means   that  $\int   d^{2}x
\sqrt{g(\mathbf{x})}  (s(\mathbf{x})+\kappa^{2})=0$  or else  that the mean
defect density is equal to the Gaussian  curvature $-\kappa^{2}$ up to some
trivial prefactor.   This charge neutrality expresses  the frustration
effect of  curvature that forces  an irreducible density of defects of
the same sign (here, point  disclinations with a negative  topological
charge). As  stated  previously, this    is  quite similar  to     the
frustration   effect  of  an   applied  uniform  magnetic   field on a
superconducting film, in which an  irreducible density of vortices  of
the same  sign is  induced. Actually,  at low  temperature and in  the
continuum limit, the dual  description of Eq.(\ref{eq:8}) in  terms of
defects leads to the following free-energy functional\cite{FT94,FT95,F80}:

\begin{equation}\label{eq:11}
F[s]=   N E_{core}  +    \frac{1}{2} Z \int d^{2}x \int  d^{2}y
(s(\mathbf{x})-f)G(\mathbf{x},\mathbf{y})(s(\mathbf{y})-f) ,
\end{equation}

where  $G(\mathbf{x},\mathbf{y})$   is the inverse   Laplacian in flat
space, \textit{i.   e.},    goes    as  the   coulombic    interaction
$log(\mid\mathbf{x}-\mathbf{y}\mid/a_{0})$   at   large distances,   and  $f
\varpropto B^{2}$.  Here too, charge  neutrality imposes that the mean
density of vortices is equal to $f$.

We have made  the above rather long  digression to illustrate Nelson's
analogy  between   glassforming  liquids    and  uniformly  frustrated
systems. Taking this analogy for granted, one may envisage as possible
minimal theoretical   models  for  glass formation  simple   uniformly
frustrated   models\cite{Note0}, either   in   the continuum   version
described above or in a lattice version, which  may be more convenient
for computer simulation  studies. The simplest of  such models is  the
two-dimensional uniformly frustrated XY model, whose Hamiltonian is

\begin{equation}\label{eq:12}
H=-J \sum_{\left\langle ij \right\rangle } \cos(\theta_{i}-\theta_{j}-A_{ij})
\end{equation}
where the  sum  is  over  distinct pairs  of  nearest-neighbor  sites,
$\theta_{i}$ is the angle of the XY  spin at site $i$,  and $A_{ij}$ is the
line integral of the  vector potential along  the bond $(ij)$; the sum
of the  $A_{ij}$'s  taken  in  a  clockwise  direction over  the bonds
forming an     elementary plaquette of  the  lattice   is fixed  to be
$\sum_{\square} A_{ij} = 2 \pi f$ for each plaquette ($f \in \left[ 0,1/2
\right]$). The frustration  parameter  $f$ is  simply  related to  the
square of  the amplitude of the uniform  applied field (see above). By
duality transformation\cite{JKKN77},  one may  alternatively focus  on
the associated defect (vortex) description: this corresponds to a $2d$
Coulomb   lattice gas in   a   uniform background  charge $-f$,  whose
Hamiltonian is just the lattice version of Eq.(\ref{eq:11}).

Simulations of the   $2d$ uniformly frustrated   XY model and   of the
associated Coulomb lattice gas model  have been performed for a  large
value  of  frustration,  $f_{*}=(3-\sqrt{5})/2$  corresponding to  the
so-called  "full  frustration",  and glassy   behavior  reminiscent of
supercooled liquids has indeed been observed\cite{KL97,LK99}.

In three dimensions,  one can generalize the  Coulomb lattice gas to a
lattice  model of interacting vortex loops,  in which vortex lines are
forced by frustration  in all directions of  space (as is required  to
model  a glassforming liquid, but no  longer corresponds to an extreme
type-II superconductor in an applied magnetic field\cite{LT93}):

\begin{equation}\label{eq:13}
H                    =                    \frac{J}{2} \sum_{i,j,\mu=x,y,z}
(s_{i\mu}-f)G(\vert\mathbf{x_{i}}-\mathbf{x_{j}}\vert)(s_{j\mu}-f)
\end{equation}

where  $s_{i\mu}$ is  the integer  vorticity  on bond $\mu$ emanating from
site $i$, which forms continuous  lines (loops or infinite lines), and
$G(\vert\mathbf{x_{i}}-\mathbf{x_{j}}\vert)$    is  the  lattice  $3d$
Coulomb    interaction        that        goes      as        $1/\vert
\mathbf{x_{i}}-\mathbf{x_{j}}   \vert$ at  large separations.   Charge
neutrality imposes that $\sum_{i,\mu}(s_{i\mu}-f)=0$.

At this point one should stress that whereas structural and thermodynamic
properties are simply obtained by studying the partition function computed by
integrating over the defect degrees of freedom, the dynamics is trickier: the
mapping from the local order parameter, which evolves with simple local
dynamics, to the associated defects may imply quite complex kinetic rules for
the defects (see section~\ref{sec:defects-effect-kinet}).

\subsection{\textbf{Competition between effective interactions: (screened)
Coulomb frustrated models}}\label{sec:textbfc-betw-effect}

Frustration describes  a competition  between  a
local tendency to  order and opposing   constraints.  It is then tempting
to  describe this  situation at a  coarse-grained   level by means  of
competing effective interactions  acting on different length scales. The
situation   seems   quite clear for       the ordering tendency:   the
corresponding interaction is  short-ranged  and the  associated  local
order parameter characterizes the  liquid LPS. Although things are not
as clear as far as the frustrating effective interaction is concerned,
there  is     a good   rationale for   taking    this  interaction  as
long-ranged. This is what we argue now.

\begin{figure}\begin{centering}
\resizebox{8cm}{!}{\includegraphics{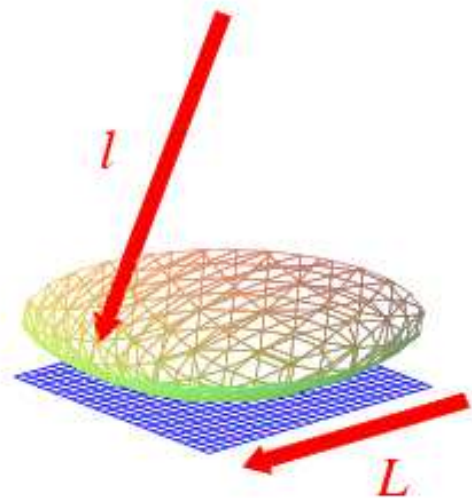}}
\caption{"Squashing" of an ideally ordered region of size $L$
from curved space to flat space; $l$ denotes the
intrinsic curvature (i. e., frustration) scale.} \label{fig:5}
\end{centering}\end{figure}
A robust feature of the effect of frustration is the fact that, at low
temperature,  forcing  the extension  of the LPS  (\textit{i. e.}, the
ideal  order) over a   region of size   $L$ induces a  super-extensive
free-energy cost, a strain energy in a continuum elastic picture, that
grows  as $L^{d+2}$  in   dimension  $d$.  A heuristic   and  somewhat
simple-minded,  argument  can  be put forward   as  follows. Imagine a
reference curved  space with  intrinsic curvature  scale  $\ell$ which is
(ideally)  ordered  at low temperature (see figure \ref{fig:5}). What  is  the effect
 of "squashing" an  ordered region of size $L$,  with $L$ much larger than
the typical  molecular size but  smaller  than $\ell$ (this  implies that
frustration is small), onto flat,  Euclidean  space? the  free-energy
density of this region in flat space can be expanded as

\begin{equation}\label{eq:14}
\Phi(L,l)     = \Phi_{0}(l)           + \Phi_{1}(l)(\frac{L}{l})   
         +
\Phi_{2}(l)(\frac{L}{l})^{2} +O((\frac{L}{l})^{3}) .
\end{equation}

The first term of the  above expression corresponds to the free-energy
gain  associated with ideal  ordering  in the  absence of  frustration
($\Phi_{0}(\ell) <  0$); the second  term is zero, $\Phi_{1}(\ell)=0$, because
one
can choose the  flat region locally  tangent to the curved space.  The
last term describes the dominant contribution to the cost generated by
frustration. As a result,  the corresponding contribution to the  free
energy  of   the region  of volume  $L^{d}$   in  flat space   goes as
$(\Phi_{2}(\ell)/\ell^{2})L^{2+d}$   for   a  space   of  dimension   $d$. This
"super-extensive" cost prevents the extension  of the LPS in Euclidean
space.

The argument can be  checked on the  models described in the two above
subsections. The $2$-dimensional case  is straightforward.  Consider the
free-energy functional described in  Eqs.(\ref{eq:6},\ref{eq:7}).  In this case,   ideal
order consists in a hexatic phase which is present in the unfrustrated
Euclidean    space and   corresponds  to  a    uniform order parameter
$\mathbf{n}(\mathbf{x})=\mathbf{n}_{0}$.  The  free energy of a region
of size $L \ll \ell=\kappa^{-1}$ with $\mathbf{n}(\mathbf{x})=\mathbf{n}_{0}$ on
the hyperbolic plane is then given at leading order in $\kappa \ell$ by

\begin{equation}\label{eq:15}
F=\frac{1}{2} \kappa^{4}   Z_{H} \vert \mathbf{n}_{0} \vert^{2} \pi \int_{0}^{L}
rdr\,  r^{2} \sim \kappa^{4}L^{4} ,
\end{equation}

where  we have used the  fact that at leading order, $g_{\mu\nu} \approx
\delta_{\mu\nu}$
and $\Gamma_{\lambda\mu}^{\nu}(\mathbf{x}) =  -\kappa^{2} \delta_{\lambda\mu}
x^{\nu}$. Here too, and  up
to possible logarithmic corrections, there  is a free-energy cost that
grows  as $L^{d+2}$ with  $d=2$.   The same reasoning  applies  to the
extreme type-II superconducting    film in a  perpendicular   magnetic
field.

In  terms of the associated defect  picture, with defects being either
point disclinations or point  vortices, the above  result can be simply
interpreted: forcing the ideal (hexatic or superconducting) order in a
region of size $L$ when frustration is present  amounts to forcing the
defect density to be zero in that region; due  to the coulombic nature
of the interactions   in  Eqs.(\ref{eq:10}) and (\ref{eq:11}) ,   this  then leads to    a
super-extensive cost that  goes as $\kappa^{4}L^{4}$  or $f^{2}L^{4}$ (plus
logarithmic corrections). A main difference between hexatic order on
the  hyperbolic plane and superconducting order  in a uniform magnetic
field is  the presence  of a  screening effect  in the  former  case at
distances larger than the intrinsic frustration (curvature) scale.

The  3-dimensional case of models   with a coupling  to a  non-Abelian
gauge background is slightly  more involved. Consider for illustrative
purpose the $SO(4)$ theory of  frustrated icosahedral order with the  local
order  parameter $\mathbf{Q}_{\ell}(\mathbf{x})$   with  $\ell=12$  (in  the
following we drop  the  index $\ell$).  Ideal order now  exists in curved
space.  Following  the  interpretation  of   Sethna\cite{Se85} and  Nelson-Widom\cite{NW84},
particle configurations in   Euclidean space that most closely  mimic
the  ideal order  are  constructed by  "rolling" a  reference polytope
$\left\lbrace 3,3,5\right\rbrace$ between adjacent  points. It is then
convenient to introduce a change of variable

\begin{equation}\label{eq:16}
\mathbf{Q}(\mathbf{x})=e^{i\kappa  \mathbf{\hat{L}}_{0\mu} x^{\mu}}
\mathbf{\tilde{Q}}(\mathbf{x})
\end{equation}

so  that the order is  now  measured relative to  a reference polytope
that has been rolled  in straight lines emanating  from the  origin in
all   directions. One  can      then try    to force  the     relation
$\mathbf{\tilde{Q}}(\mathbf{x})=\mathbf{Q}_{0}$  in a region of linear
size $L$   around  the origin,  with   $\mathbf{Q}_{0}$ minimizing the
potential   part of   the  free-energy  functional   (\textit{i.  e.},
$\frac{1}{2} \tau \mid\mathbf{Q}(\mathbf{x})\mid^{2} +  O(\mathbf{Q}^{3})$). The
free-energy cost is associated with the gradient term $\frac{1}{2} Z \int
d^{3}x \mid D_{\mu}\mathbf{Q}(\mathbf{x})\mid^{2}$.

By appying the Baker-Campbell-Hausdorff formula

\begin{equation}\label{eq:17}
e^{A}e^{B}  = e^{A + B   + \frac{1}{2} \left[ A,B \right]+\frac{1}{12}
(\left[   A,\left[    A,B  \right]\right]   +    \left[  B,\left[  B,A
\right]\right] )+...}
\end{equation}

to

\begin{equation}\label{eq:18}
e^{i\kappa  \mathbf{C}_{\mu}(\mathbf{x})} =  e^{-i\kappa    
\mathbf{\hat{L}}_{0\mu}
x^{\mu}} e^{i\kappa \sum_{\nu} \mathbf{\hat{L}}_{0\nu} x^{\nu}} ,
\end{equation}

where  we have made  explicit the sums over  repeated indices to avoid
ambiguity, one obtains that

\begin{equation}\label{eq:19}
i\kappa   \mathbf{C}_{\mu}(\mathbf{x})=   i\kappa \sum_{\nu \neq \mu} 
\mathbf{\hat{L}}_{0\nu}
x^{\nu}  -\frac{1}{2}   (i\kappa)^{2} \sum_{\nu}  
\left[\mathbf{\hat{L}}_{0\mu}  ,
\mathbf{\hat{L}}_{0\nu} \right]x^{\nu}x^{\mu} + O(\kappa^{3}x^{^{3}}).
\end{equation}

The covariant derivative can then be rewritten
\begin{equation}\label{eq:20}
D_{\mu}\mathbf{Q}(\mathbf{x})=e^{i\kappa \sum_{\nu}  \mathbf{\hat{L}}_{0\nu} 
x^{\nu}}
(\partial_{\mu} + \mathbf{\Gamma_{\mu}(\mathbf{x})})
\mathbf{\tilde{Q}}(\mathbf{x}) ,
\end{equation}

with the "connection" $\mathbf{\Gamma_{\mu}(\mathbf{x})}$ now defined as

\begin{equation}\label{eq:21}
\mathbf{\Gamma_{\mu}(\mathbf{x})} = i\kappa
\partial_{\mu}\mathbf{C}_{\mu}(\mathbf{x}) = -\frac{1}{2} (i\kappa)^{2}
\sum_{\nu} \left[\mathbf{\hat{L}}_{0\mu} , \mathbf{\hat{L}}_{0\nu}
\right]x^{\nu} + O(\kappa^{3}x^{^{2}}).
\end{equation}

It is easy to check that the first term of the  expression is equal to
the covariant curl,  $-(i/2)\mathbf{F_{\mu\nu}}x^{\nu}$, which is non-zero due
to  the non-Abelian property  of the  gauge  group $SO(4)$; it is also
equal to the curvature tensor and in the  present case is proportional
to the gaussian curvature $\kappa^{2}$.

When $\mathbf{\tilde{Q}}(\mathbf{x})=\mathbf{Q}_{0}$,  the    gradient
term can  be  expressed at leading  order in  $\kappa \ell$ (with the Einstein
summation convention again) as

\begin{eqnarray}\label{eq:22}
F_{cost}(L)     &=&         \frac{1}{2}         Z \int_{(L)}      d^{3}x \mid
\mathbf{\Gamma_{\mu}(\mathbf{x})}    \mathbf{Q}_{0}\mid^{2}    \nonumber\\&    =&
\frac{1}{8}                                                    Z \kappa^{4}
(\mathbf{\hat{L}}_{\mu\nu}\mathbf{Q}_{0})^{*}(\mathbf{\hat{L}}_{\mu\nu'}\mathbf{
Q}_{0})
\int_{(L)} d^{3}x x^{\nu} x^{\nu'},
\end{eqnarray}

where  a star denotes complex conjugation and where we  have   used  the   commutation relations  of   the  $SO(4)$
generators, $\left[ \mathbf{\hat{L}}_{0\mu},\mathbf{\hat{L}}_{0\nu}\right]
=i\mathbf{\hat{L}}_{\mu\nu}$. It   is trivial  to  derive  from  the above
equation that the free-energy   cost grows as $\kappa^{4}L^{5}$  with  some
irrelevant prefactor\cite{Note1}.

If, as argued   above, frustration indeed generates  a super-extensive
free-energy cost   opposing the  extension  of  the ideal  order,  the
associated  effective interaction should  be long-ranged: the integral
of a coulombic interaction $\propto 1/r^{D-2}$ over a  region of size $L$ in
dimension $d$ precisely leads to an energy  going as $L^{d+2}$. Such a
Coulomb-like  dependence is  a  very  general feature of  interactions
between defects, be they  vertices in superconductors, dislocations in
crystals,   or disclinations  in   bond-orientationally ordered  phases
\cite{K89}. Therefore, whereas the extension of the liquid LPS is more
conveniently   formulated  in  terms   of   a local  order  parameter,
frustration is  more  conveniently   formulated in terms    of  defect
densities, akin to local "disorder"  parameters. A full-fledged  gauge
field theory   could   possibly handle this    duality  (see above and
Ref. \cite{K89}). However, a much simpler picture is provided by working with
a   single    variable $\mathbf{S}(\mathbf{x})$    for   all effective
interactions. This  leads   to (possibly  screened) Coulomb-frustrated
models. In   a   continuum  description,  the  associated   free-energy
functional is given by

\begin{eqnarray}\label{eq:23}
F[\mathbf{S}]&=&\int   d^{3}x  \lbrace            \frac{1}{2}          Z \mid
\partial_{\mu}\mathbf{S}(\mathbf{x})\mid^{2} +
\frac{1}{2} \tau \mid \mathbf{S}(\mathbf{x})\mid^{2}+ O(\mathbf{S}^{3}) \rbrace 
\nonumber\\& -& \frac{1}{2} K \int d^{3}x \int d^{3}y \mathbf{S}(\mathbf{x})
G(\vert\mathbf{x}-\mathbf{y}\vert) \mathbf{S}(\mathbf{y}),
\end{eqnarray}

where      $K>0$     is     a  measure        of   the     frustration
strength.  $G(\vert\mathbf{x}\vert)$ is  the Coulomb interaction  that
behaves  as $1/\vert\mathbf{x}\vert$  in   $3d$, possibly screened  at
distances much  larger    than  an    intrinsic  frustration    length
$\ell$.  (Typically, such a  screening  leads to  a Yukawa pair potential,
$\exp(-(\vert\mathbf{x}\vert/\ell))/\vert\mathbf{x}\vert$.) In the   above
equation, we have  been cavalier with the tensorial character
of the variables and the interactions.

One may as well consider a lattice version

\begin{equation}\label{eq:24}
H=-J \sum_{\left\langle ij  \right\rangle  }
\mathbf{S}_{i}\mathbf{.}\mathbf{S}_{j} +
\frac{K}{2}\sum_{i,j}  \mathbf{S}_{i}   G(\mathbf{x}_{i}-\mathbf{x}_{j})
\mathbf{S}_{j}
\end{equation}

which  is more convenient for  computer   simulations. The models  are
chosen  such that in the  absence  of frustration ($K=0$),  there is a
continuous  or  weakly first-order transition  to   an ideally ordered
phase at a temperature of order $J$.

In the corrugated-space picture\cite{N83,SM99,GSM82}, one can interpret the
above Coulomb-frustrated  models as describing the competition between
the  extension of  the   ideal order in regions  which   look like the
reference  space, \textit{i. e.},  regions with an effective curvature
that allows  tiling by the  LPS, and the  repulsion between regions of
alike  curvature, which   enforces the  constraint  that the   average
curvature must be zero (see also Ref.\cite{GKT97}).

Models similar  to those  described   here have  been used   in  quite
different contexts. In   diblock  copolymers formed by   two  mutually
incompatible polymer chains  attached to   each other, the   repulsive
short-range forces between the two types of components tends to induce
phase separation of  the melt, but  total segregation is forbidden  by
the    covalent bonds      that   link    the    subchains    together
\cite{L80,OK86,BF90}.  A       microphase    separation
transition occurs instead at   low enough temperature and  the  system
then forms periodically modulated phases, such as lamellar, hexagonal,
and  cubic phases.  Similarly,  self-assembly in  water-oil-surfactant
mixtures results from the competition  between the tendency of oil and
water  to phase separate and  the stoichiometric constraints generated
by the  presence of surfactant molecules,  constraints that act as the
electroneutrality    condition  in   a system   of   charged particles
\cite{WCS92,DC94,WWSW02,S83}. The same kind of  physics also arises in
a very different situation: stripe formation in doped antiferromagnets
like  cuprates has  been   ascribed to  a frustrated electronic  phase
separation, by which a  strong local  tendency of  the holes  to phase
separate    into  a  hole-rich   "metallic"   phase and   a  hole-poor
antiferromagnetic phase is  prohibited  by  the  long-range  coulombic
repulsion between the holes\cite{EK93}.

The  behavior of Coulomb-frustrated    models  has been  studied  both
through scaling arguments and  by  computer simulations.  The  results
will    be   described  in      sections \ref{sec:phen-scal-appr}  and
\ref{sec:comp-simul-simple}.

\section{Avoided criticality}\label{sec:avoided-criticality}

It   seems a little  distressing that  no unique, unchallenged minimal
theoretical model   for frustration in   liquids has  been  derived so
far. However, the situation  is not as   bad as it looks, since  there
appears to  be universal or to  the least robust properties  shared by
the models  described in the   above section. The most  important such
property         is              that       of        \textit{"avoided
criticality"}\cite{KKNT95,CEKNT96,NRKC99}.  This notion  expresses the
fact that  the  critical  point  which    exists in  the  absence   of
frustration and  separates    a disordered (liquid)  from   an ideally
ordered  phase  disappears as  soon as    an  infinitesimal amount  of
frustration is introduced.

\begin{figure}\begin{centering}
\resizebox{11cm}{!}{\includegraphics{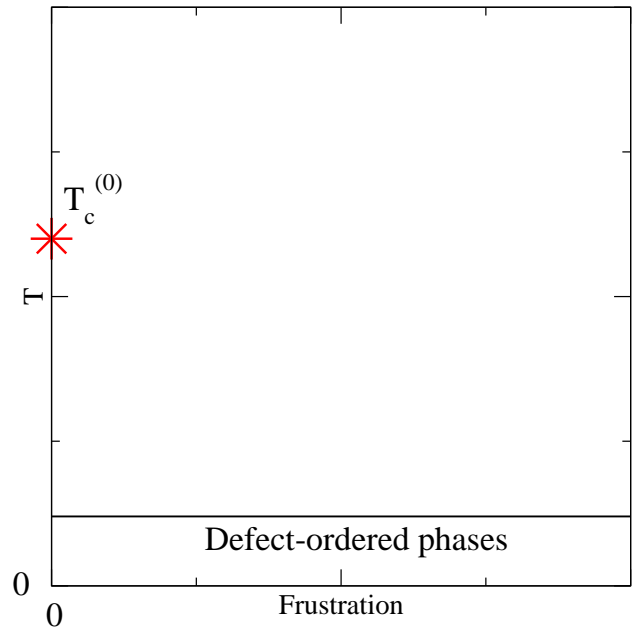}}
\caption{Schematic temperature-frustration diagram illustrating
the property of avoided criticality (a genuine discontinuity when frustration goes to zero).} \label{fig:6}
\end{centering}\end{figure}

Let us try  to make  this more explicit.   Long-range ideal order   is
forbidden  in the presence of frustration,  but  other types of order,
which  we generically describe as "defect  ordered  phases", are still
possible.  By defect-ordered  phases,  we  mean  phases  in which  the
defects that  break the ideal  order themselves  arrange in a periodic
array. This includes: Frank-Kasper-like  phases in materials with local
icosahedral  order,  in which  the  disclination lines form  a regular
network;  the  Abrikosov flux  lattice  in type-II superconductors, in
which the vortices form  a lattice; and in Coulomb-frustrated systems,
phases with modulated order, in  which domain walls arrange in regular
patterns  leading  for instance  to  lamellar or   stripe  phases. The
phenomenon of avoided phase transition   then means that the limit  to
zero  frustration of the  possible transitions to those defect ordered
phases  is at a temperature  $T_{DO}(frustration \to 0^{+})$ significantly lower than
$T_{c}^{(0)}$,  that of the critical point  in  zero frustration. This
discontinuity, which    is illustrated  in figure \ref{fig:6},  is a  genuine
non-perturbative effect due   to the strong  fluctuations generated by
frustration.   A similar   phenomenon  occurs if    the transition  at
$T_{c}^{(0)}$ is weakly   first-order\cite{N04}.    (For a  strong    first-order
transition,  the  transition may also  be  avoided, but the associated
phenomenology is different\cite{JKS05}.)  It is worth stressing
that this  property  of avoided   criticality  has been overlooked  in
earlier studies of geometric frustration\cite{N02,SN85}.

\subsection{\textbf{Evidence for avoided criticality in the frustrated models}}\label{sec:textbf-avoid-crit}

We first discuss  the cases   where  frustration is described   either
through  a  coupling   to a non-Abelian   gauge background   or by  an
effective Coulomb interaction. Proving   the existence of  an  avoided
critical point then implies several steps.

The first step is to  show that the minimum  of the interaction kernel
$\Delta(\vert  \mathbf{x}-\mathbf{y}\vert)$,  or in  Fourier  space $\Delta(q)$,
which  appears in the  quadratic part  of  the free energy functional,
\textit{i. e.}, in

\begin{equation}\label{eq:25}
F[\mathbf{S}]=          \frac{1}{2} \int          \frac{d^{d}q}{(2\pi)^{d}}
\mathbf{S}(\mathbf{-q})\Delta(q)\mathbf{S}(\mathbf{q}),
\end{equation}

or a similar expression for the models defined in (Eq.(\ref{eq:3})), occurs at
a
non-zero wave  vector with modulus $q_{0}$.  In the Coulomb frustrated
models (Eq.(\ref{eq:23})), the interaction kernel can be expressed as

\begin{equation}\label{eq:26}
\Delta(q)=q^{2}+(4\pi K)/q^{2}+constant,
\end{equation}
where  we have   used  that  the Fourier  transform  of   the  Coulomb
interaction   is equal  to  $4\pi/q^{2}$; the   minimum then occurs  for
$q_{0}=(4\pi  K)^{1/4}$  (here  and in the   following  we have  set for
convenience $Z\equiv1$). In the screened Coulomb case,  the above result is
modified to

\begin{equation}\label{eq:27}
\Delta(q)= q^{2} + \ell^{-2} + (4\pi K)/(q^{2} + \ell^{-2})+constant,
\end{equation}

with  the minimum now   at $q_{0}=(4\pi K-\ell^{-2})^{1/2}$.   For physical
reasons (see section  \ref{sec:textbfc-betw-effect}), one  expects the
coupling parameter $K$ to be related to the inverse of the frustration
length, $K \sim \ell^{-\varpi}$,  so  that the  minimum   is again at a   nonzero
wave-vector whenever $0<\varpi\leq4$.

\begin{figure}\begin{centering}
\resizebox{11cm}{!}{\includegraphics{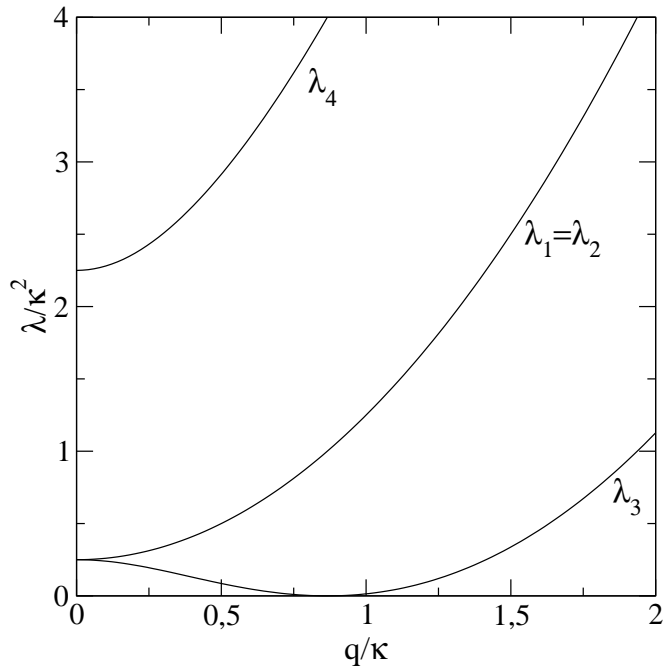}}
\caption{Eigenvalues of the quadratic part of the Hamiltonian
describing $O(4)$   spins with non-Abelian  $SO(4)$  couplings  in $3$
dimensions  in   the  continuum limit  (see  text).  There   are four eigenvalues, $\lambda_1=\lambda_2=\mu-6+\kappa^2+q^2$,
$\lambda_{3,4}=\mu-6+2\kappa^2+q^2\mp\kappa \sqrt{\kappa^2+4q^2}$, where $\kappa$ is the frustration
strength and $\mu$ is  chosen so that the  minimum  value of  the lowest
eigenvalue is  equal to zero.  From  top to  bottom, $\lambda_3,\lambda_1=\lambda_2$ and
$\lambda_4$  divided by  $\kappa^2$, plotted as   a function of the dimensionless
wave   vector ($q/\kappa$).     The minimum   in $\lambda_3$   occurs   on  a shell
$q_0/\kappa=\sqrt{3}/2$.}
\label{fig:7}
\end{centering}\end{figure}

For the   models with    a  coupling   to  a gauge    background,  the
demonstration  is more  involved.  The interaction  kernel  in Fourier
space is not only a  function of $q$ but also  an operator or a matrix
that should be diagonalized.  It can be  checked in the $SO(4)$ theory
of  Nelson and coworkers   (see  Eqs.(\ref{eq:1}) and (\ref{eq:2}))  that  all  the  eigenvalues
$\lambda_{mm'}^{\ell}(q)$ of the kernel, associated with a given representation
of   dimension $(\ell +1)^{2}$ of  $SO(4)$,  have a single   minimum at a
non-zero  value $q_{0},_{mm'}^{\ell}$  that  goes  as $\kappa$  for small 
$\kappa$ (where $\kappa$ is   the    coupling strength  associated   with 
geometric frustration): see  for instance figure $6$ of  Ref.\cite{SN85}.
In figure \ref{fig:7} we also plot as another example the eigenvalues for the simpler case
of $O(4)$ spins with $SO(4)$ couplings in $3$ dimensions \cite{N04}.

The second step consists in studying the models in the limit where the
number of components $N$ of the local order  parameter is large. Exact
analytic results can be obtained when $N \to \infty$\cite{CEKNT96} (this corresponds to the
spherical  limit for  hard spins and  to  the gaussian variational  or
Hartree  approximation   for   soft spins). A     phase transition (of
second-order in this case) occurs whenever

\begin{equation}\label{eq:28}
-\tau= \int\frac{d^{d}q}{(2\pi)^{d}}\frac{1}{\lambda(q)-\lambda(q_{0})},
\end{equation}

where $\tau$ is defined  in Eqs.(\ref{eq:3},\ref{eq:23})  and $\lambda(q)$ is  the smallest  of the
eigenvalues  of the interaction  kernel.  A similar expression  can be
derived for a lattice, hard-spin version\cite{CEKNT96}. It is easy to check
that the  integral on the right-hand side  is dominated by wavevectors
whose modulus $q$ is near $q_{0}$; in this domain, the denominator can
be approximated by   a quadratic term proportional  to $(q-q_{0})^{2}$
whereas the integration measure is proportional to $dq q_{0}^{(d-1)}$,
so  that    when  $q_{0}\neq0$ the  integral  strongly    diverges in all
dimensions  $d$.  Therefore,  no ordering  transition   is possible at
finite $\tau$, \textit{i. e.}, at non-zero temperature.

When the model is considered on a  lattice and not in continuum space,
symmetry-breaking effects are  generated that may induce a  transition
at a   non-zero temperature. However,       in dimension $d<4$    this
temperature is of the order of the  magnitude of the symmetry-breaking
terms,  quite different  and    lower than the   critical  temperature
$T_{c}^{(0)}$ in  the unfrustrated system\cite{CEKNT96}.  This leads to the
phase diagram illustrated  in  figure~\ref{fig:6}.  $1/N$ corrections have  also
been  computed and  shown   to leave  unchanged  the  avoidance of the
critical point\cite{NRKC99}.

As a third step,  one can go beyond the  large $N$ limit (note however
that the dimension  of the main  component of the order parameter  for
local icosahedral order in Nelson's  $SO(4)$ theory is equal to $169$,
quite  a large value) and evaluate  the magnitude of the fluctuations
about the putative ordered ground state. This can be done in the usual
and somewhat  heuristic   way by studying  the  quadratic fluctuations
only. If the number of components $N$  of the local order parameter is
large enough, one always find  "transverse" fluctuations whose size is
given  by   the same   integral  as that  on  the   right hand side of
Eq.(\ref{eq:28}). Because of the  strong  divergence of the  integral,
the ground  state is always  unstable with respect to such transverse
fluctuations:   as  a  result,  no   ordering  takes  place  at finite
temperature as soon   as  frustration   is present\cite{N04,NRKC99}. In    the  Coulomb
frustrated models,  this  is what occurs for   $N>2$. For  $N=2$,  the
divergence of the integral is  only finite-temperature logarithmic, which suggests that a
phase with quasi-long range order is possible
\cite{NRKC99}.

The case $N=1$ (Ising), as more generally the  case of all models with
a discrete instead of a continuous symmetry, is also special. As first
shown   by Brazovskii\cite{B75}   and  further  supported by  computer
simulations\cite{VT98,GTV01b},  there is  a first-order  transition to
modulated phases; this first-order transition is induced by the strong
fluctuations  associated  with   frustration  (indeed the   mean-field
approximation predicts   a  second-order  phase  transition)   and its
temperature $T_{DO}$ goes continuously to $T_{c}^{(0)}$ as frustration
goes  to zero.   However, the limit   of  stability (spinodal) of  the
disordered phase is  depressed to  zero  temperature and  the critical
point  at $T_{c}^{(0)}$ remains  the  only finite-temperature  critical
point in   the  temperature-frustration diagram.   The  transition  at
$T_{DO}$ being   first-order, it should  be  possible  in principle to
supercool the disordered   (liquid) phase to lower  temperature.  This
will    be       discussed     in    more     detail     in    section
\ref{sec:comp-simul-simple}.

The  above reasoning can   be made more rigorous  with  the help of  a
generalized  Mermin-Wagner    inequality\cite{N04,N01};  it  shows  that
whenever the integral

\begin{equation}\label{eq:29}
\int\frac{d^{d}q}{(2\pi)^{d}}\frac{1}{\lambda(\vert  
\mathbf{q}+\mathbf{q_{0}}
\vert)-\lambda(\vert \mathbf{q}-\mathbf{q_{0}} \vert)-2\lambda(q)}
\end{equation}

diverges,  ordering at wave vector   $\mathbf{q_{0}}$ is inhibited at
finite temperature.

Finally, some results are also    available concerning the   uniformly
frustrated   spin   models.     In    the   $2$-dimensional    version
(Eqs.  (\ref{eq:8}),~(\ref{eq:9})),  there appears  to  be an  avoided
critical point, just  like in the  models described above. In this
case, the defect-ordered phases are the Abrikosov vortex lattice and the associated hexatic
vortex phase, and vanishing frustration means vanishing applied magnetic field. As shown by
D. Fisher\cite{F80}  on  the  basis  of  the  continuum Coulomb  gas description
(Eq.(\ref{eq:11})) within the KTNHY theory  of $2$-dimensional melting
(for a review, see \cite{Nelson83}), the transitions to the  defect-ordered phases occur, in the limit of vanishing frustration, at temperatures more than an order  of magnitude lower than the Kosterlitz-Thouless
transition\cite{KT72,V72} that takes place at  zero frustration/zero magnetic field; this has also been confirmed  by computer simulations\cite{FT94,FT95,GT00}.
The  resulting phase  diagram,   as   illustrated  by  figure  1    in
Ref. \cite{F80}, is then similar to that shown here in figure  ~\ref{fig:6}.

Again,  the physics  of the transitions   in the  presence and  in the
absence of frustration  is  very different.  The  melting  transitions
describe the unbinding of the  dislocations and disclinations that are
present in the low temperature lattice or hexatic phases formed by the
vortices induced by the applied magnetic field. On the other hand, the
Kosterlitz-Thouless  transition at $T_{c}^{(0)}$   describes the unbinding of  a neutral
assembly of vortices    of both positive  and   negative  signs.   The
$3$-dimensional version describing  an  extreme type-II superconductor
in  an  applied  field does  not  seem to   have  an  avoided critical
point\cite{T99,OT05}, but  the case of an  isotropic
frustration discussed above in section   \ref{sec:textbfc-betw-effect}
has not been studied.

\subsection{\textbf{Consequences of an avoided critical point}}

In dimension $d>1$  an Ising  ferromagnet,  when placed  in a  uniform
external field, has of   course some sort  of avoided  criticality: an
infinitesimal field destroys the  critical fluctuations beyond a given
length scale. However, the phenomenon is trivial. The external magnetic
field induces  a non-zero magnetization at  all temperatures,  so that
the system  is always  ferromagnetic and  does  not undergo any  phase
transition.

Quite different  is the situation  in a ferromagnet with $N$-component
vector  spins  (and $O(N)$  symmetry)  in  the presence  of a quenched
random field (with zero mean). For dimensions between $2$ and $4$, the
pure  system (no random  field) possesses a  critical point associated
with the usual    transition from  paramagnetism  to   ferromagnetism,
whereas the random-field model,  provided $N\geqslant3$, has no  phase
transition     at      all   and       stays paramagnetic     at   all
temperatures\cite{AP83}; indeed, the lower critical dimension is $d=2$
in the  former case and $d=4$  in the latter. Interesting, non-trivial
scaling properties emerge  in such a  case for temperatures  equal and
less   than $T_{c}^{(0)}$,  the     critical    point of  the     pure
system\cite{AP83}.   This  is to  such  phenomenon that we  refer when
exploiting   the    property  of    avoided   criticality  shown    by
frustration-based models of glassforming liquids.

On   general grounds, one  expects that   the  presence of an "avoided
critical point" leads to several diverging length scales, according to
whether  one approaches the   critical point  at  zero frustration  by
varying  the temperature, or one approaches  the critical point or the
ideally  ordered  phase by decreasing   the frustration  to zero.  A
phenomenological scaling  approach will be  developed and
discussed in the next section.

\section{Phenomenological scaling approach of glassforming liquids:
frustration-limited domain theory}\label{sec:phen-scal-appr}

The  above developments suggest   two main conclusions concerning  the
description of glassforming liquids\cite{KKNT95}:

1) Frustration naturally leads to collective (or cooperative) behavior
on   a  mesoscopic  scale,   a  feature that  we     stressed as being
characteristic       of    the       phenomenology     (see    section
\ref{sec:phen-what-there}).   The  collective property  comes from the
phase growth (in which the liquid LPS spreads in space) induced by the
proximity of an avoided critical point,  whereas the limitation on the
scale over such a growth can take  place results from frustration that
aborts the phase transition and leads to "avoided criticality".

2) The  relevant temperature about which  one  can organize  a scaling
description   of the  viscous slowing     down and other    collective
properties  of supercooled liquids  is  that  of the avoided  critical
point in the associated unfrustrated  system. This, of course, is only
meaningful if the critical   point is narrowly avoided,  which implies
that the frustration characterizing  the liquid under consideration is
small enough. Note also that in a liquid the temperature $T_{c}^{(0)}$
marks a crossover, not a  true transition, so that  it is not expected
to be sharply defined.

\subsection{Heuristic scaling arguments} \label{sec:heur-scal-argum}

As a first step towards a scaling analysis, one can simply incorporate
the physics   of  frustration-induced   avoided  criticality   in  the
consideration of aborted nucleation of the  ideal ordered phase in the
liquid phase.  At temperatures  sufficiently below $T_{c}^{(0)}$,  the
free  energy of a   single domain of the   ideal phase in a disordered
liquid environment, when its characteristic linear size is $L$, can be
written as\cite{KKNT95,KZKFK94}

\begin{equation}\label{eq:30}
F(L,T)=\sigma(T) L^{\theta}-\phi(T) L^{3}+s(T) L^5.
\end{equation}

The   first two terms are   those   commonly encountered in  classical
nucleation  theory:  the  first one  represents  the  free-energy cost
associated with  building  a  domain of   one  phase in  another;  for
simplicity, and  in the absence  of more  specific information, we set
$\theta=2$, which corresponds to a surface tension term associated with the
interface between the two phases\cite{Note2} . The  second term is the
bulk free-energy difference between the two  phases and corresponds to
a gain  in free energy.   If $\xi_{0}$ is the  correlation length of the
unfrustrated sytem, the surface tension  $\sigma$ scales as $\xi_0^{-2}$  and
the ordering bulk free-energy density $\phi$ as $\xi_0^{-3}$. The last term
represents   the   strain free-energy  resulting    from the effect of
frustration; as discussed in section \ref{sec:textbfc-betw-effect}, it
is generically expected to grow super-extensively with system size, as
$L^{5}$ in $d=3$.   The above equation assumes  that  the size $L$  is
large  compared to the typical molecular  length, so that a continuum,
thermodynamic-like description can be  used, but is small compared  to
the intrinsic characteristic scale of   frustration, so that one  does
not   have account  to for  the  screening   effect that modifies  the
super-extensive growth of the strain free energy at long distances.

Because  frustration in a  liquid should be the  same  in any point of
space, one actually expects,  instead  of the  formation of a   single
domain,  the breaking up of the  liquid into a  collection of domains,
or,  to use  more     pictorial terms, a  "mosaic"\cite{KTW89}   or  a
``pathwork''\cite{FHS88} of domains.  Those domains are separated from
each  other by "interfaces", i.  e.,  regions in which the ideal order
is broken and a higher  concentration of defects, such as disclination
lines, is present.   The size of the  domains and their further growth
when lowering  the  temperature is limited   by frustration.  Assuming
that the inter-domain interactions are sufficiently weak to be treated
by a  mean-field approach, the   free-energy density, $\Phi(L,T)$, for  a
domain of  linear size $L$ in a  mosaic of frustration-limited domains
(with randomly oriented    local  order parameters)  can be   directly
derived  from    the free-energy of  a    single  domain  given above,
namely\cite{TKV00},

\begin{equation}\label{eq:31}
\Phi(L,T)=\frac{\sigma(T)}{L}-\phi(T)+s(T)L^2,
\end{equation}

where  $\sigma$, $\phi$, and $s$ are  only trivially renormalized with respect
to their values for an isolated domain.

The   typical domain size    $L^*$   is  obtained by  minimizing   the
free-energy  density   given  above,  and  this   leads  to $L^{*}(T)\sim
(\sigma/s)^{1/3}$;  the  strain coefficient  $s$   is,  however,  \textit{a
priori}  unknown. The scaling analysis  thus relies on  (at least) two
supermolecular structural   lengths,  the correlation length    of the
unfrustrated system,   $\xi_{0}(T)$,   and the  typical    domain  size,
$L^*(T)$. As   one  decreases the    temperature below  $T_{c}^{(0)}$,
$\xi_{0}(T)$  \textit{decreases}   whereas   $L^*(T)$ is  expected    to
\textit{increase}. Scaling only makes sense below a temperature $T_{1}
\lesssim T_{c}^{(0)}$ at which $\xi_{0}(T)$ and $L^*(T)$ are comparable to each
other  and   both  larger  than the    typical  molecular length. Such
conditions may be only marginally satisfied in real liquids.

It   is  important to  stress that  the  mosaic of frustration-limited
domains is an equilibrium feature (in the  supercooled liquid phase it
is of  course metastable  with  respect to  the actual  crystal):  the
domains, as well as the overall pattern they form,  are not frozen but
continuously   changing due  to  molecular  motion;  their statistical
properties,  however, depend only on   the thermodynamic state of  the
liquid.

In such  a  mosaic of domains, $\alpha$   relaxation and flow  must involve
restructuring of domains. In the simplest scaling picture, the typical
free-energy barrier   to  be overcome  for  such  a  restructuring, $\Delta
E^{*}(T)$,  scales as  $\sigma(T)L^{*}(T)^{2}$,   \textit{i.  e.}, involves
motion,   creation, or destruction of a   domain wall. A more detailed
argument for such a description has  been given by Stillinger in terms
of a "tear and repair" mechanism\cite{FHS88}. Scaling about a narrowly
avoided critical point then implies that both $\sigma(T)\propto\xi_{0}(T)^{-2}$ and
$L^{*}(T)$   have   a  power-law behavior    in  $(T_{c}^{(0)}-T)$ for
temperatures less than $T_{c}^{(0)}$.  (Recall    that at and    above
$T_{c}^{(0)}$ there are no frustration-limited  domains.) As a result,
the typical free-energy barrier for $\alpha$ relaxation scales as

\begin{equation}\label{eq:32}
\Delta E^{*}(T) \sim (T_{c}^{(0)}-T)^{\psi}.
\end{equation}

In a naive, mean-field-like  picture, $\sigma(T)$ and $L^{*}(T)$  vanish as
one  approaches $T_{c}^{(0)}$    from   below in    an  analytic  way,
$\sigma(T)\propto(T_{c}^{(0)}-T)$ and     $L^{*}(T)\propto(T_{c}^{(0)}-T)$, so
that one
obtains $\psi=3$.  In a more  elaborate, but still heuristic\cite{Note2} description
,
it has been found  that  $L^{*}(T)$ goes as   $K^{-1/2}\xi_{0}(T)^{-1}$,
where $K$ is the  relative  amplitude of the frustrating   interaction\cite{KKNT95}
(see     Eqs.(\ref{eq:23},~\ref{eq:24})):   the  typical  domain       size   then grows  as
$(T_{c}^{(0)}-T)^{\nu}$, where $\nu$ is the correlation length exponent of
the unfrustrated system,  as temperature decreases; it  also increases
as frustration decreases and diverges in the limit of zero frustration
as expected   in a scenario  of avoided  criticality.  Taking $\nu\approx2/3$,
which is characteristic of  ordinary critical phenomena in the absence
of  quenched disorder in  $3$ dimensions, one arrives  in this case at
$\psi\approx8/3$.

The above   estimates concern   the  collective contribution   to  the
free-energy   barrier   for   $\alpha$  relaxation,   that    due  to   the
frustration-limited     domains   which   vanish    around  and  above
$T_{c}^{(0)}$.  One must also  account for  the molecular contribution
which characterizes "ordinary" liquid  behavior (typically, above  the
melting temperature $T_{m}$). The simplest description that provides a
reasonable fit to the experimental  data in the ordinary liquid  phase
without introducing any  spurious  singularities is a  plain Arrhenius
formula,

\begin{equation}\label{eq:33}
\tau_{0}(T)=\tau_{\infty}\exp(\frac{E_{\infty}}{k_BT}),
\end{equation}

where $\tau_{\infty}$  and  $E_{\infty}$ are  species-specific molecular 
constants
characteristic of the high-temperature liquid.

The main   prediction  of this   scaling  approach  is  thus that  the
$\alpha$-relaxation   time  and   the  viscosity    have  an activated-like
dependence on temperature,  with a \textit{crossover}  from molecular,
ordinary  liquid   behavior  to  collective, domain-dominated behavior
around   $T^{*}\thickapprox T_{c}^{(0)}$, a   crossover  that leads   to   a
change from
Arrhenius-like to superArrhenius $T$-dependence,

\begin{equation}\label{eq:34}
\tau_{\alpha}(T)=\tau_{\infty}\exp(\frac{E(T)}{k_BT}),
\end{equation}

with $E(T)=E_{\infty}+\Delta E^{*}(T)$.  The superArrhenius  contribution to the
activation free energy is described by a universal power law,

\begin{equation}\label{eq:35}
\begin{array}{cllc}
\Delta E^{*}(T)&=&   0   &   {\rm  for   }\,\,    T>T_{c}^{(0)},\nonumber\\
&=&Bk_BT_{c}^{(0)}\left(1-\frac{T}{T_{c}^{(0)}}\right)^{\psi}&  {\rm  for
}\,\, T<T_{c}^{(0)}.
\end{array}
\end{equation}

\begin{figure}\begin{centering}
\resizebox{10cm}{!}{\includegraphics{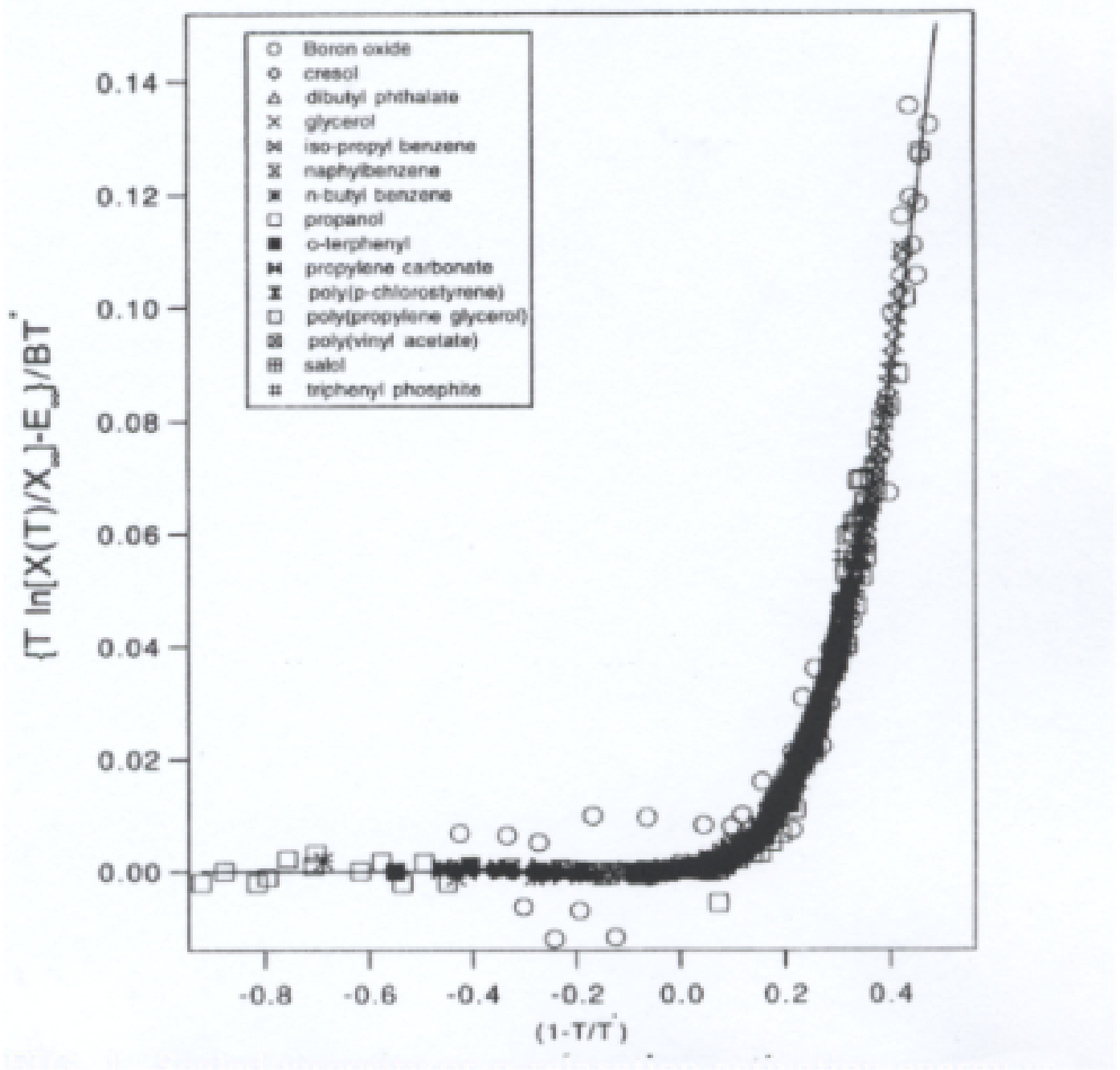}}
\caption{Scaling of the collective component of the effective
activation (free) energy, $\Delta E^*(T)$, versus reduced
temperature $(T^*-T)/T^*$ measured from the crossover temperature $T^*$
(or equivalently, the avoided critical temperature
$T_c^{(0)}$). The solid line is the $8/3$ power law discussed
in the text. Reprint from Ref.\cite{KTZK96}.} \label{fig:8}
\end{centering}\end{figure}

This prediction fits well the extant experimental data on glassforming
liquids with $\psi$ in the range from $7/3$ to  $3$ (the best overall fit
being for $\psi=8/3$)\cite{KKNT95,KTZK96}, a range compatible with  the above estimates. This
is shown in  figure \ref{fig:8}. In the fitting  procedure, $T_{c}^{(0)}$ is  an
adjustable parameter: for virtually all liquids  studied, its value is
found slightly above  the  melting  point $T_{m}$; this  indeed  makes
sense since one  expects the stability of  the ideal ordered phase  in
the unfrustrated system at $T_{c}^{(0)}$ to be higher than that of the
actual crystal in the frustrated liquid\cite{S84,N04}.

The parameter   $B$  in Eq.(\ref{eq:35}) is  a  measure of  the  departure from
Arrhenius behavior, hence of ``fragility''. From the scaling analysis,
we  obtain  that $B$ increases as   frustration decreases (it  goes as
$K^{-1}$ for   Coulomb frustrated models). One   can  thus associate a
large fragility with a small frustration. This  is compatible with the
view developed  in section~\ref{sec:geom-frustr-simple}   that atomic  glassformers   made of
spherically  symmetric particles are not  fragile,  or in the standard
terminology  are rather  "strong",    despite  the weakness    of  the
inter-particle interactions: frustration in those systems characterized
by local icosahedral order is indeed strong.

\subsection{Further dynamic and static scaling analysis}\label{sec:furth-dynam-stat}

It is  tempting  to  push further  the  scaling  analysis in order  to
address other aspects of the phenomenology of supercooled liquids.

The equilibrium distribution of domain sizes  in the mosaic, $\rho(L,T)$,
at temperatures   below  $T_{1} \lesssim T_{c}^{(0)}$  can  be  obtained from
$\rho(L,T)\propto\exp\left[-(\Phi(L,T)-\Phi_{*}(T))L^3/k_BT)\right]$,  where
$\Phi(L,T)$
is the free-energy density given in Eq.(\ref{eq:31}) and $\Phi_{*}(T)$ its value
at
the minimum. It can be cast in a scaled form as\cite{TKV00,VTK00}

\begin{equation}\label{eq:36}
\rho(L,T)\propto        \exp\lbrace-\gamma(T)[\kappa(\frac{L}{L^*})^2      
-\frac{3}{2}(\frac{L}{L^*})^3+\frac{1}{2}(\frac{L}{L^*})^5]\rbrace,
\end{equation}

where the entire temperature  dependence is contained in the parameter
$\gamma(T)\propto  B(T_{c}^{(0)}/T)(1-T/T_{c}^{(0)})^{\psi}$;   $\kappa$ is   a number  of
order  1 introduced to   allow for small   domain-shape  effects at  a
mean-field level\cite{VTK00} (it   should not  be  confused with   the frustration
parameter  $\kappa$   used   in sections   \ref{sec:geom-frustr-simple} and
\ref{sec:stat-mech-frustr}).

By generalizing the simple dynamic scaling  arguments presented in the
above subsection, one can also evaluate the collective contribution to
the activation free energy, $\Delta E(L,T)$, for a domain of size $L$; this
latter takes the following scaled form\cite{TKV00,VTK00}:

\begin{equation}\label{eq:37}
\frac{\Delta E(L,T)}{k_BT}=b\gamma(T)[(\frac{L}{L^*})^2-m(\frac{L}{L^*})^5],
\end{equation}

where $b$ and $m$ are numbers of order 1, with $bm<1/2$. The last term
in $L^{5}$ has been included to account for the  fact that building or
moving a domain wall costs a free energy of  order $\sigma L^{2}$ but leads
to  a reduction in strain  free energy. The  full effective activation
free energy is then obtained  by adding the molecular,  Arrhenius-like
contribution, $E_\infty /k_BT$, to the above expression.

\begin{figure}\begin{centering}
\resizebox{11cm}{!}{\includegraphics{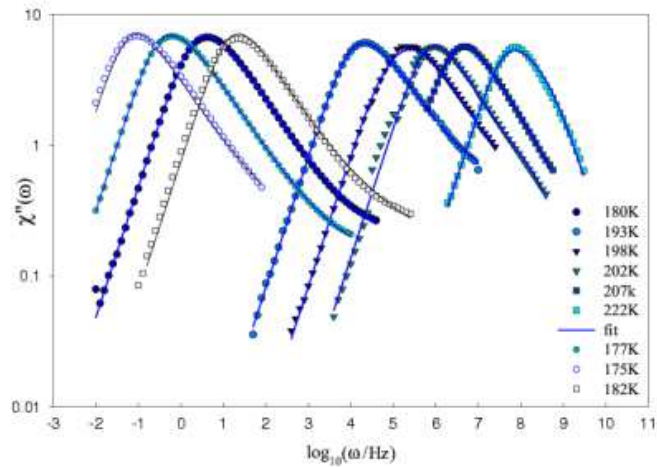}}
\caption{Frequency-dependent dielectric susceptibility of the
fragile  glassformer  m-fluoroanilin  for several  temperatures   on a
log-log plot: comparison  between  experimental data ($T_g=172K$)  and
frustration-limited  domain  scaling    theory  (data  and   fits   by
C. Alba-Simionesco,  see also Ref.\cite{ALBA01}).  The parameters used
for the    fit  (see text)   are   ~$T^*=T_c^{(0)}=258K$, $E_\infty=1587K$,
$\tau_\infty=10^{-12}s$, $\kappa=0.62$, $m=0.29$, $b=1.04$, $B=364$. The exposant $\psi$ is fixed to $8/3$.}
\label{fig:9}
\end{centering}\end{figure}

Below the temperature $T_{1}\lesssim T_{c}^{(0)}$ at  which the liquid breaks
up into supermolecular domains,  the dynamics are naturally  described
as  \textit{heterogeneous}. For instance, the nonexponential character
of the dielectric  relaxation can be  simply modeled by assuming  that
molecular  reorientations  are  completed within a  single  domain, in
which they are  dynamically coupled  to the   order variable, and   by
assuming  that  the relaxation within  each  domain  is exponential in
time\cite{R02}. (Of course, there may  be systems such as polymers for
which the bare  "molecular" relaxation is already non-exponential; the
collective   behavior  associated  with  the  mosaic  of  domains then
increases the non-exponential  character.) This leads to the following
expression    for         the    normalized   dielectric    relaxation
function\cite{TKV00,VTK00}:

\begin{equation}\label{eq:38}
f_\alpha(t)=\int_0^\infty      dL  L^2
\rho(L,T)\exp\left[-\frac{t}{\tau_{\infty}
}\exp\left(-\frac{E_\infty +\Delta E(L,T)}{k_BT}\right)\right],
\end{equation}

with   $\rho(L,T$)    and  $\Delta  E(L,T)$ given   in   Eqs.(\ref{eq:36}) and
(\ref{eq:37}).   The above expression  compare  well with experimental
data  on glassforming   liquids\cite{VTK00}.   This is illustrated  in
figure \ref{fig:9} where the imaginary part of the frequency-dependent
dielectric    susceptibility   of   the    fragile   glassformer   $\alpha$
m-fluoroanilin is compared  to the Fourier transform of $df_{\alpha}(T)/dt$
obtained from  Eqs.(\ref{eq:36}), (\ref{eq:37}) and (\ref{eq:38}): the
main features of the $\alpha$  relaxation are well  reproduced over a large
frequency  range (up   to    13  decades)  and a  large    temperature
domain\cite{ALBA01}. (The approximate  scaling plot proposed  by Nagel
and           coworkers\cite{DWNWC90}          is       also      well
reproduced\cite{TKV00,VTK00}.)   One  of   course  recovers  for   the
temperature dependence of the  peak  frequency the power-law  behavior
described in Eq.(\ref{eq:35}).   The  number of adjustable  parameters
strongly  restricts the predictive power of  the scaling approach, but
it should be stressed that those parameters are
\textit{independent  of temperature}.  From  those  parameters one can
estimate the typical domain size, which is found to  be around 5 to 10
molecular diameters at $T_{g}$,  an  estimate that is compatible  with
experimentally determined sizes (see section
\ref{sec:geom-frustr-simple}); with additional input\cite{TK95}, one can also describe the
decoupling     between  translational  diffusion  and  reorientational
relaxation\cite{VTK00}.

A few additional points are worth stressing: (i) in this approach, the
heterogeneous   character of the   $\alpha$ relaxation  in the  supercooled
liquid regime  is  a  consequence of a   \textit{structural} property,
namely,    the    break-up   of  the      liquid  into a   mosaic   of
frustration-limited  domains; this  is  to  be contrasted  with  other
descriptions that  focus     on purely kinetic  mechanisms     (see the
review article\cite{RS03}). (ii)  The superArrhenius growth of the relaxation
time with decreasing temperature is accompanied  by only only a modest
stretching of the relaxation  functions (compare for instance with the
quite  different behavior  of magnetic  systems in the   presence of a
random  field\cite{HF87};  in  the present  scaling  description, this
results   from  the  decreasing polydispersity    of   the domains  as
temperature decreases.

Finally, scaling  about an avoided critical  point gives a qualitative
account  of the  rapid decrease  of  the excess (or "configurational")
entropy  observed in supercooled liquids  as  one approaches the glass
transition. Within this approach,  there are indeed  two sources  of entropy decrease  below
$T_{c}^{(0)}$ as one  lowers  the temperature: the  first one   is the
increase  of ordering that  should  take place inside  all the domains\cite{KT98}, the second  is associated with the  growth  of the typical
domain  size  and the resulting decreasing  weight  of the interfaces
with high   concentration  of defects.  This  latter   effect will be further
discussed  in section \ref{sec:conn-other-appr},  in
connection  with the description  of glassforming  liquids in terms of
the complexity of their free-energy landscape and  the large number of
metastable states.  Just as the $\alpha$-relaxation  time and the viscosity
rapidly increase  with  decreasing  temperature with  no  need  for  a
low-temperature  singularity, the frustration-limited domain  approach
predicts a rapid  decrease of the  entropy with  no  requirement for a
low-temperature Kauzman catastrophe.

The  above phenomenological  scaling   analysis illustrates  how   the
concept  of  frustration   coupled   with  the   property of   avoided
criticality  can provide a  framework for describing the phenomenology
of glassforming liquids. However, whether or  not the assumptions that
underlie     the   scaling  analysis   evolve    rigorously  from  the
statistical-mechanical         models   discussed   in         section
\ref{sec:stat-mech-frustr}  remains  a    fundamental      theoretical
uncertainty.  This point has not  yet been resolved, but some progress
has      been    achieved   by     means    of     computer-simulation
studies of  simple frustrated models. This  is what
we discuss next.

\section{Computer simulations of simple systems: Coulomb frustrated lattice
models}\label{sec:comp-simul-simple}

The  Coulomb    frustrated    spin  models   described     in  section
\ref{sec:textbfc-betw-effect}   seem   to be   the  simplest   systems
compatible      with   the   frustration-based    approach   of  glass
formation. Setting aside  the issue of  their microscopic derivation,
it is possible to check the relevance  of the scenario presented above
by  investigating  numerically   such  models. Extensive   Monte-Carlo
simulation studies\cite{GTV01b,GTV02,GTV02b} have been performed for models on a $3$-dimensional
cubic lattice with Hamiltonian (see also Eq.(\ref{eq:24}) )

\begin{equation}\label{eq:39}
H=-J \sum_{\left\langle ij   \right\rangle }
\mathbf{S}_{i}\mathbf{.}\mathbf{S}_{j} +
\frac{K}{2}\sum_{i\neq                                                   j}
\frac{\mathbf{S}_{i}\mathbf{.}\mathbf{S}_{j}}{\vert\mathbf{x}_{i}-\mathbf{x}_{j
}\vert}
\end{equation}

for a variety of  spin variables, mostly  with discrete symmetry: Ising
model, ${\bf  S}_i     = ±1$,   $q$-state   clock    model,   ${\bf
S}_i=(\cos(2\pi\theta_i/q),\sin(2\pi \theta_i/q))$, with $\theta_i$ the orientation    of
the planar spin and $q=5$ and $q=11$. We also include here new results
obtained   for the XY model, \textit{i.    e.},  for continuous planar
spins.
\begin{figure}\begin{centering}
\resizebox{11cm}{!}{\includegraphics{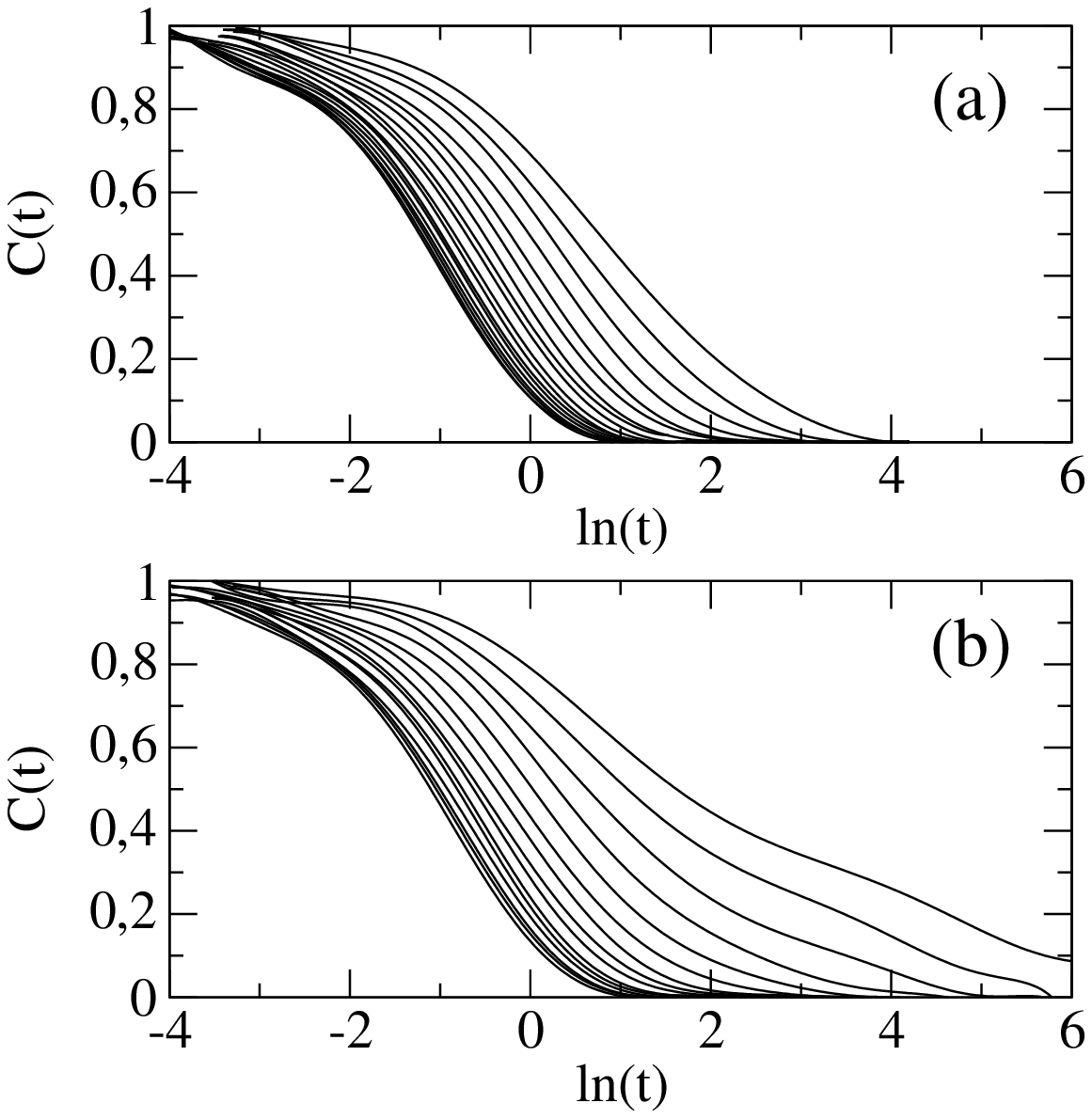}}\\
\resizebox{11cm}{!}{\includegraphics{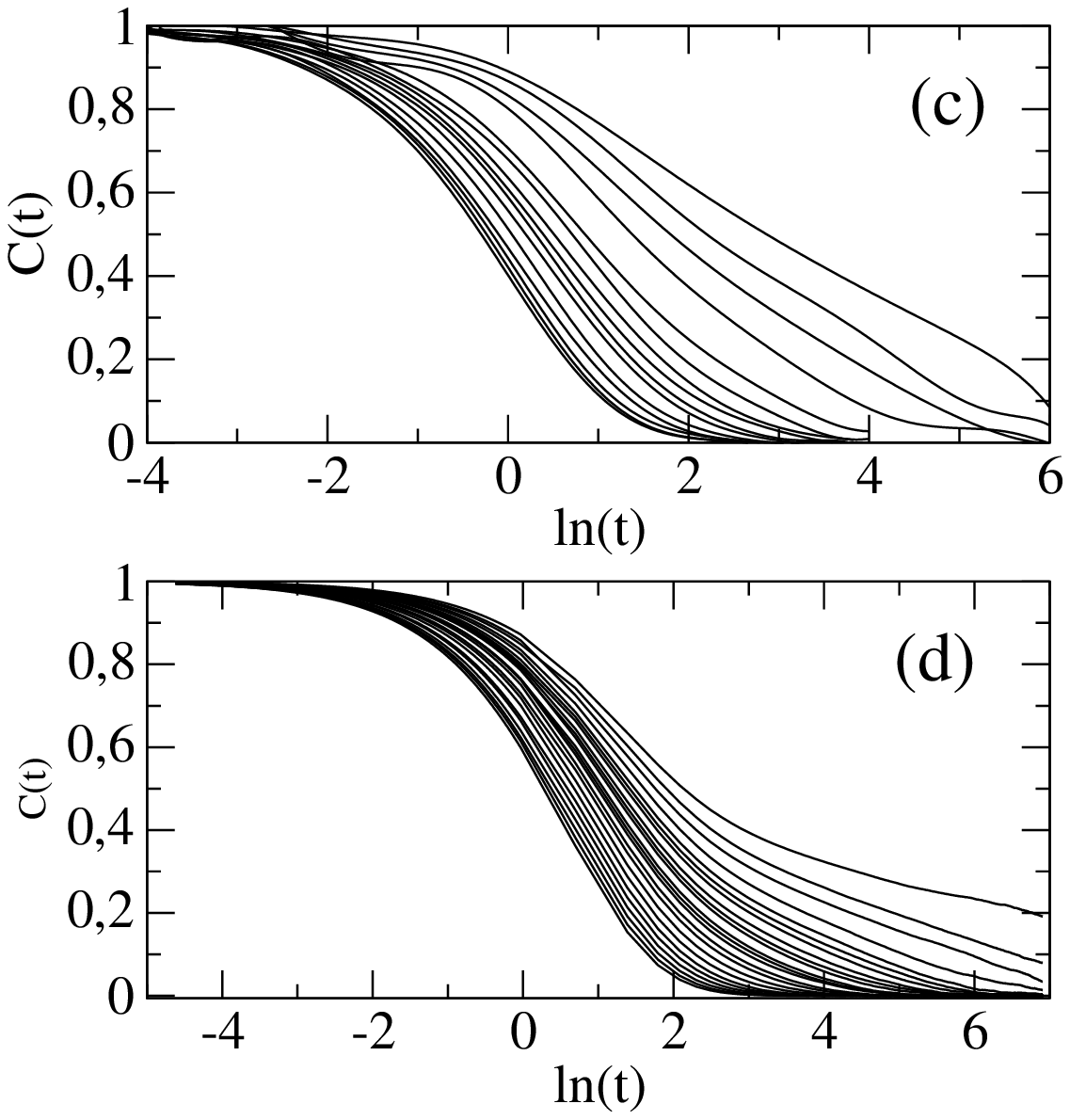}}
\caption{Spin-spin correlation function  $C(t)$ versus $ln(t)$ for
several temperatures both above and below $T_c^{(0)}$ (in
each panel, curves from left to right are for
decreasing temperatures): $5$-state clock model for
$Q=0.1$ (a) and $Q=0.00625$ (b); $11$-state clock model for
$Q=0.001$ (c); and XY model for $Q=0.005$ (d).} \label{fig:10}
\end{centering}\end{figure}

In  the  absence of  frustration  ($Q=0$), the models  have a critical
point   at a    temperature    $T_c^{(0)}$,  below  which  they    are
ferromagnetic. The long-range, coulombic interaction requires that the
total magnetization of the system be zero  in order to ensure a proper
thermodynamic  limit. Therefore,  long-range   ferromagnetic order  is
prohibited  at  all $T's$  for  any nonzero   value of the frustration
parameter $Q/J$. (In what follows,  we take $J=1$.) The  models studied
(XY and  discrete orientations) display only  a "weak" form of avoided
criticality.   They can still  form   ordered or quasi-ordered  phases
characterized by modulated   patterns (e.g., lamellar  phases) below a
temperature $T_{DO}(Q)$ that goes non-analytically but continuously to
$T_c^{(0)}$ as $Q$  goes  to  zero. However, the  transition   between
paramagnetic and  modulated phases is  first-order\cite{VT98,GTV01b}. Following
Brazovskii\cite{B75}, this  result can be  interpreted on the basis of
the self-consistent Hartree approximation which predicts the occurrence
of a fluctuation-induced first-order  transition, a transition with no
nearby low-$T$ spinodal. As a   result, $T_c^{(0)}$ remains the   only
nearby  (and  avoided) critical   point in the frustration-temperature
diagram. All  the   simulations  have  been  made in   the  disordered
(paramagnetic) phase. Since the  transition to the modulated phases at
$T_{DO}(Q)$  is first-order,  one  could  in principle  supercool  the
paramagnetic phase to  lower temperatures; however,  it has been found
that the  lattice sizes accessible in  practice are too small to allow
for  a  proper   supercooling below  $T_{DO}$\cite{GTV02,GTV02b}.  This,   unfortunately,
restricts the   domain  of temperature over  which  one  can study the
slowing down of the dynamics.  (Out-of-equilibrium dynamics and  aging
phenomena have not been considered.)

The  relaxation to equilibrium   of the  frustrated  systems has  been
studied      via   the    Monte     Carlo     algorithm    (Metropolis
rule), time being the  number of sweeps per  spin. The
dynamical    quantity  that  has  been monitored    is the equilibrium
spin-spin self  correlation function, $C(t)=(1/N)\sum_i<{\bf S}_i(t'){\bf
S}_i(t'+t)>$, where the bracket denotes the thermal average and $N$ is
the  total number of lattice  sites; cubic lattices  of size $16^3$ to
$22^3$  with  periodic boundary   conditions  have been  used  and the
Coulomb interaction has been handled via Ewald sums\cite{GTV02,GTV02b}.

The  evolution of $C(t)$  with $T$  is illustrated in figure \ref{fig:10}
for the  $5$-state    clock ($Q=0.1$   and  $Q=0.00625$),  the   Ising
($Q=0.001$), and     the XY  ($Q=0.005$)  models.  At      the highest
frustration, illustrated   by figure \ref{fig:10}a,  the decay of  $C(t)$
appears to proceed  in a single step   at all temperatures.  For lower
frustrations,  a  $2-$step  decay  develops  as  $T$  is  lowered (see
figure \ref{fig:10}b-d).  At   high temperature, typically   above   the
critical point  of  the corresponding unfrustrated  model ($T_c^{(0)}\simeq
2.1,  4.51, 2.32$  for  the $5-$state clock,   the  Ising, and the  XY
models,  respectively),   and for all   frustrations,  the  whole time
dependence of  $C(t)$   is  well  fitted  by a   simple   exponential,
$\exp(-t/\tau_0(T))$.   At  low  temperature,  below $T^{(0)}_c$,  it  is
impossible  to  describe   the entire  decay  of  $C(t)$  by  a single
exponential; the  emerging   second  step of the   relaxation   can be
described by a stretched exponential, $\exp(-(t/ \tau_{KWW}(T))^{\beta(T)})$,
while  the first  step can still  be  fitted by  a simple exponential,
$\exp(-t/ \tau_0(T))$\cite{GTV02,GTV02b}.  This  non-exponential behavior
and   emergence  of  a  two-step  decay,   both  of  which become more
pronounced  as $T$  is  decreased, are  typical of fragile supercooled
liquids. One may however notice that  the timescale separation between
the two relaxation steps  that can be achieved  in the  simulations is
not sufficient to    observe the development of   a  true  plateau  at
intermediate times.
\begin{figure}\begin{centering}
\resizebox{10cm}{!}{\includegraphics{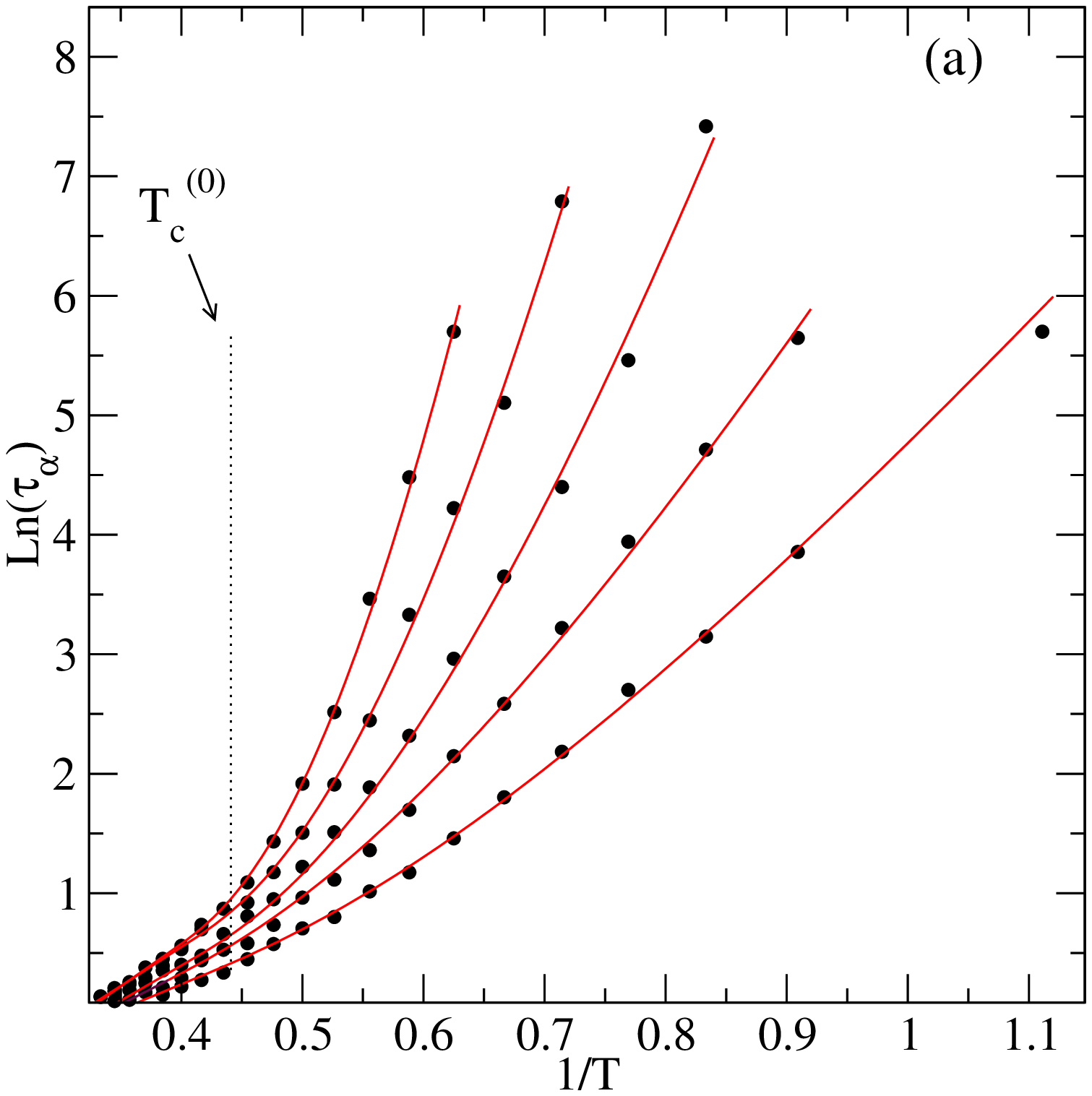}}\\
\resizebox{10cm}{!}{\includegraphics{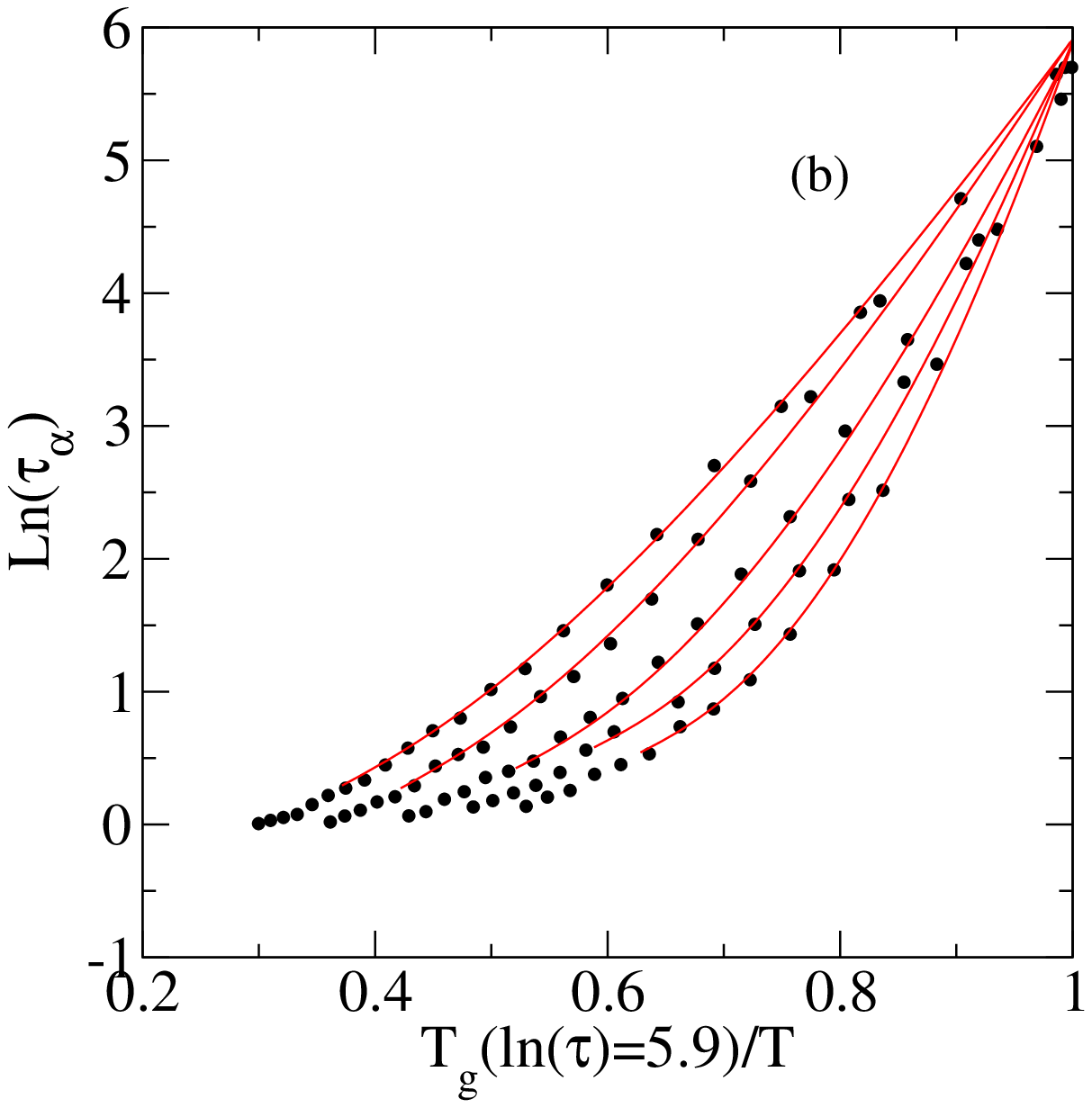}}

\caption{Arrhenius plot of $\tau_\alpha(T)$ for the $5$-state clock model and
 various frustrations  (from  left   to   right in  (a)  and from right to left in (b),  $Q=0.00625,
 0.00125,0.025,0.05,0.1$).  In (b) the   curves are plotted versus the
 reduced   inverse  temperature $T_{g}(\ln(\tau)=5.9)/T$,   where  $T_{g}(\ln(\tau)=5.9)$ is the
 temperature at which the   maximum relaxation time, $\ln(\tau_\alpha)=5.9$,  is
 attained.} \label{fig:11}
\end{centering}\end{figure}

The slowing down  of the relaxation  as one lowers  the temperature is
visible in  figure ~\ref{fig:11}.   The temperature  dependence of  the
relaxation times characteristic of the short- and long-time behaviors,
$\tau_0(T)$  and $\tau_\alpha  (T)$ respectively,     has been investigated    in
Refs.\cite{GTV02,GTV02b} .  Over the range of temperature studied, the
dependence  of  $\tau_0(T)$   is essentially    Arrhenius-like,  $\tau_0(T)\simeq
\tau_{0,\infty}\exp(E_\infty  /T)$,  whereas the dependence  of   $\tau_\alpha(T) $ shows a
marked  deviation from     Arrhenius behavior   below   some crossover
temperature in the vicinity of $T^{(0)}_c$.

The  crossover from Arrhenius to  super-Arrhenius behavior of $\tau_\alpha$ is
shown  in figure \ref{fig:11} for the   $5-$state clock model for  several
different frustrations. A similar trend is observed  for the Ising and
XY models. Several points are worth noting: (i) such a super-Arrhenius
behavior   is   typical  of the   viscous     slowing down of  fragile
glassforming liquids, the more  fragile  a liquid the  more pronounced
the  super-Arrhenius character;    (ii) the crossover   occurs in  the
vicinity of the  critical point of the  unfrustrated system; (iii) the
departure from Arrhenius  behavior, \textit{i.e.}, the  ``fragility'',
becomes more marked as frustration decreases.

As shown  in   Refs.\cite{GTV02}  and \cite{GTV02b}, the   temperature
dependence of the $\alpha$-relaxation time  is compatible with the behavior
predicted  by     the   frustration-limited   domain    approach  (see
Eq. (\ref{eq:35})).   However,   because  of the   limited domain   of
temperature, hence the limited  range of relaxation times, one  should
not put too much emphasis on the fit: for instance,  the data are also
compatible     with     a    Vogel-Fulcher     formula,     $\tau_\alpha=\tau_{0}
exp(DT_{0}/(T-T_{0}))$\cite{GTV01}; on the other hand, they cannot  be fitted by a
power-law dependence.

\begin{figure}\begin{centering}
\resizebox{10cm}{!}{\includegraphics{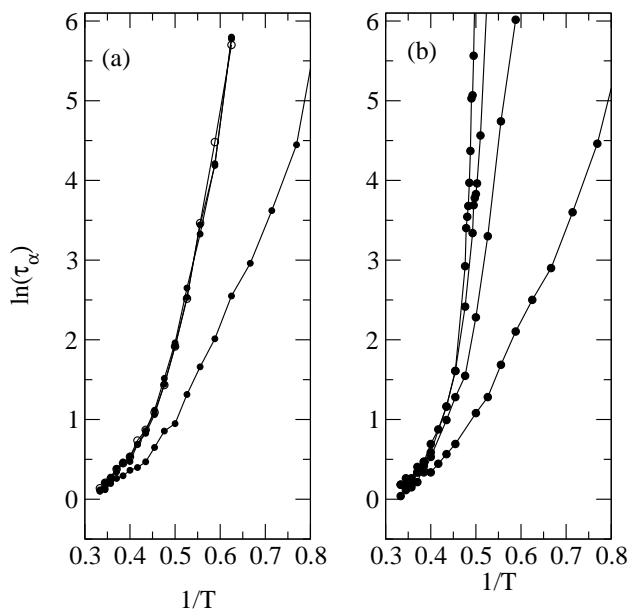}}
\caption{Arrhenius plot of  $\tau_\alpha$   for the  $5-$state   clock  model
and for   lattice   sizes $L=5,10,15,20$  (from  right  to  left): (a)
$Q=0.00625$ (no difference is seen between the $3$ larger sizes)  and  (b) $Q=0$ (standard critical slowing down).} \label{fig:12}
\end{centering}\end{figure}

Although  no direct determination  of a domain or heterogeneity length
has been  attempted in those  studies,  interesting insight  has  been
provided  by investigating the finite-size   effect on the  relaxation
time.   As illustrated  in   figure \ref{fig:12}a, the  behavior is  quite
different than that of standard  critical slowing down (shown here for
the   unfrustrated  system, figure \ref{fig:12}b): in  the  presence of
frustration, the slowing down of the relaxation does not come with the
rapid   growth of an  associated length, which is compatible with the
experimental observations on glassforming liquids.

The computer simulation studies of Coulomb frustrated spin models thus
provide some   evidence that such    models display many of  the  main
characteristic features of the  phenomenology of fragile  glassforming
liquids.   It  is  however fair   to  say (i)   that the  analogy with
supercooled  liquids   is limited  by  the  range  of relaxation times
accessible in simulations  (due to the presence in  those models of  a
first-order     transition     to   modulated,   defect-ordered phases) and
(ii) that  the mechanisms by  which frustration induces a slowing down
of the relaxation remain at the level of speculations.

\section{Connection to other approaches of the glass transition: complex
free-energy landscape and kinetic constraints}\label{sec:conn-other-appr}

So far, we have stressed how frustration provides a physical mechanism
for generating cooperative behavior on  a mesoscopic scale, a  feature
that  seems to be   at the very root  of  the phenomenology of fragile
glassforming liquids. We have focused on a generic property induced by
frustration,  namely avoided criticality, and   on the possibility  of
developing a  scaling approach of  supercooled liquids exploiting this
property.   However,  starting  from the   same statistical mechanical
description based on frustration,   other routes are possible.  It  is
fruitful  to make the connection with  other, maybe more common, lines
of thought concerning the emergence of glassy dynamics: the "landscape
paradigm",  which relates  the slowing  down  of  the  dynamics to the
presence of a  large number of  long-lived metastable states, and  the
approaches in terms of  "kinetically   constrained models", in   which
glassiness  results from constraints on  the effective kinetics of the
system. Whether or  not one description  is better than the others and
to what degree the various  approaches are compatible one with another
are still, we believe, open questions.

\subsection{Long-lived metastable states and complex free-energy landscape} 

On general grounds, one expects  frustration to give  rise to a  large
degeneracy of low-lying  energy configurations\cite{G77,D04,SM99,N02}.
A first  insight into the presence of  long-lived metastable states in
supercooled  liquids can be   gained  from a heuristic argument   that
follows   from the phenomenological     scaling approach discussed  in
section~\ref{sec:phen-scal-appr}.  If   indeed  below some temperature
$T_{c}^{(0)}$    a glassforming liquid    splits   into  a  mosaic  of
frustration-limited  domains,  if  such  domains or  small  groups  of
domains are characterized by a limited number of long-lived metastable
states whose exploration via thermal activation  gives rise to the $\alpha$
relaxation,  and if those  domains  are  only weakly interacting  (see
section \ref{sec:furth-dynam-stat}), then  the system  as  a whole  is
characterized  by  an exponentially  large (in  system size) number of
long-lived metastable states: those  states are obtained  by combining
the metastable states of the roughly independent mesoscopic regions,
which  does  lead to an exponential  total  number. Since  the typical
domain  size is predicted  to  increase when  either $T$ decreases (at
constant frustration) or  frustration decreases (at constant $T$), the
number    of metastable states is expected    to  decrease when $T$ or
frustration decreases.  The  logarithm   of the number of   long-lived
metastable  states can be used to  define a "configurational entropy".
A non-zero "configurational entropy"  thus emerges below the crossover
temperature  $T_{c}^{(0)}$ associated with  the avoided critical point
and decreases as temperature further decreases.

How   can one check  the above  property?   The difficulty is twofold.
First, a  precise definition of  a "long-lived metastable  state" must
involve a time scale and, as a consequence, cannot $\textit{a priori}$
be treated within a purely   static calculation; secondly, one has  to
find an operational  way  to  compute the  number of  such  metastable
states.  In the absence  of a better  solution\cite{BK01}, a way to to
handle  the first aspect is to   assume that the long-lived metastable
states, if  indeed present in  the system,  have an infinite lifetime:
such   an  approximation,   which neglects   the   thermally activated
processes  that allow the system  to escape from the metastable states
and are the source of the  $\alpha$ relaxation, has an intrinsic mean-field
nature. The other aspect of the  problem has found an elegant solution
proposed  by  Monasson\cite{M95}:  it  relies on  introducing  a  weak
coupling to a pinning field that prevents the system from sampling all
the metastable  states  (hence breaking ergodicity)  and  handling the
average  over the pinning field by  means of the  replica method, much
used in theories of spin glasses and  other systems in the presence of
quenched disorder\cite{Y98,Note6}.

When     applied       to      the       models       introduced    in
section~\ref{sec:avoided-criticality}  to   describe   frustration  in
liquids,  this  "$\textit{replica  mean-field approach}$"  proceeds as
follows\cite{ScWo00,WSW01}.  Consider a frustrated model described (in
the  continuum  limit) by  the Hamiltonian  or  free-energy functional
$F[\chi]$  ($\chi(\textbf{x})$   stands for    $\textbf{Q}(\textbf{x})$   in
Eq.(\ref{eq:3})  , $\textbf{S}(\textbf{x})$ in Eq.(\ref{eq:23}),  etc)
and add a  weak coupling to  a pinning  or "ergodicity-breaking" field
$\sigma(\textbf{x})$.  The associated modified  partition function is given
by

\begin{eqnarray}\label{eq:40}
\tilde{\cal Z}_{\lambda}\left[\sigma \right] &=& \int  \mathcal{D\chi}
\exp\left[-\beta F[\chi] -
\frac{\lambda}{2}\int d^{d}x\left( \sigma(\mathbf{x}) -
\chi(\mathbf{x})\right)^{2}\right] \nonumber\\&
=& exp\left( -\beta \tilde{\cal F}_{\lambda}\left[\sigma \right]\right) ,
\end{eqnarray}

with  $\beta=1/(k_BT)$ and $\lambda >0$. If  $F[\chi]$  possesses low-lying minima,
$\tilde{\cal   Z}_{\lambda}\left[\mathbf{\sigma}  \right]$ will be  large  at low
temperature when $\sigma(\textbf{x})$  is close to  a  configuration of the
field $\chi(\textbf{x})$ corresponding to one of those minima. A weighted
average of the free energy in the various metastable configurations is
then obtained as

\begin{equation}\label{eq:41}
\bar{\tilde{\cal F}}_{\lambda}= \int  \mathcal{D\sigma} \tilde{\cal F}_{\lambda}\left[
\sigma \right] \frac{ e^{-\beta \tilde{\cal F}_{\lambda}\left[
\sigma \right]}}{ \int  \mathcal{D\sigma}e^{-\beta
\tilde{\cal F}_{\lambda}\left[
\sigma \right]} }.
\end{equation}

When $\lambda$ is equal to zero, $\tilde{\cal F}_{\lambda}\left[ \sigma
\right]$ reduces to the thermodynamic free energy of the system, ${\cal F}=-k_BT\ln (\int 
\mathcal{D\chi} exp\left[-\beta F[\chi] \right])$ and so does
$\bar{\tilde{\cal F}}_{\lambda=0}$. The behavior as $\lambda\to 0^{+}$ may
however be subtle: if the number of metastable states (low-lying minima) is
exponentially large in system size, the approach $\lambda\to 0^{+}$ is
non-perturbative and a discontinuity $\bar{\tilde{\cal F}}_{\lambda \to
0^{+}}-\bar{\tilde{\cal F}}_{\lambda=0}$ is present; the magnitude of the
discontinuity is precisely, up to a factor $T$, the "configurational entropy",
defined as the logarithm of the number of metastable states. Again, this can
only happen in a mean-field description, the system in the limit
$\lambda\to 0^{+}$ being stuck forever in one of the exponentially many
metastable states.

The replica trick can now be used to compute the average over the pinning field
in Eq.(\ref{eq:41}). Defining

\begin{equation}\label{eq:42}
\beta {\cal F}(m) = -\frac{1}{m} \ln {\cal Z}(m) = -\frac{1}{m} lim_{\lambda\to 0^{+}}\int 
\mathcal{D\sigma} \tilde{\cal Z}_\lambda\left[\sigma \right]^{m} ,
\end{equation}

one obtains after straightforward manipulations
that $\bar{\tilde{\cal F}}_{\lambda\to 0^+}=\partial(m{\cal F}(m))/\partial m\vert_{m=1}$ and
the configurational entropy, $S_{c}=\beta\left( \bar{\tilde{\cal F}}_{\lambda\to 0^+} -
{\cal F}\right)$ is equal to $\beta \partial {\cal F}(m)/\partial m\vert_{m=1}$, whereas the
replicated partition function ${\cal Z}(m)$ can be written as

\begin{equation}\label{eq:43}
\fl {\cal Z}(m) = lim_{\lambda\to 0^{+}} \int
\prod_{a=1}^{m}\mathcal{D\chi}_{a} \exp[-\beta \sum_{a=1}^{m}F[\chi_{a}]
 +\frac{\lambda}{2m} \sum_{a,b=1}^{m} \int d^{d}x
\chi_{a}(\mathbf{x}) \chi_{b}(\mathbf{x})] .
\end{equation}

The actual computation of ${\cal Z}(m)$ requires of course the introduction of
approximations, mean-field-like approximations as stressed above. The first
such calculation on a frustrated model has been performed by Schmalian and
Wolynes\cite{ScWo00} on the Coulomb frustrated model (Eq.(\ref{eq:23})) with an approximation known
as the self-consistent screening approximation (SCSA)\cite{AJBa74,Note4} They find that below a temperature 
$T_{A}$ an extensive configurational entropy, \textit{i.~e.}, an exponentially
large number of metastable states appear; the entropy decreases as $T$ decreases
and becomes negative below a temperature $T_{K}$. 

These results call  for   several  comments:  (1) they    confirm  the
heuristic picture described above  and, as expected, $T_{A}$ is always
found below  $T_{c}^{(0)}$, the  critical point  of the  model without
frustration;  one also  obtains    that the  configurational   entropy
decreases at constant  temperature when frustration decreases. (2) The
scenario is similar to  that characterizing generalized  spin glasses,
such  as the  $p$-spin of the  Potts glass  models, in the  mean-field
limit\cite{Note6}.  In those models, the  upper temperature  $T_{A}$ is associated
with   a dynamic  singularity   described by  the  ideal mode-coupling
approach   whereas the lower   temperature  $T_{K}$  corresponds to  a
\textit{bona fide} thermodynamic  transition to a spin-glass phase. It
can actually  be  shown that  the temperature $T_{A}$   in the Coulomb
frustrated model is also associated with a dynamic singularity similar
to that  of the  ideal  mode-coupling description\cite{GKTV02}.  One must
however recall that  this occurs within mean-field  approximations and
that such a dynamic singularity should vanish in an exact description.
The  analogy with generalized mean-field  spin glasses is the basis of
the random first-order transition  theory of glass formation developed
by  Wolynes and  his coworkers\cite{KTW89,XW00,XW01,LW01,LW03}.  We shall come
back to this theory below.

The  same property,  namely the  appearance of  a complex  free-energy
landscape with an exponentially large number of metastable states\cite{Note7}, has
been found  in     other systems related  to    the  frustration-based
description  of glassforming liquids, systems  in which the minimum of
the interaction kernel   in Fourier space  occurs  at a non-zero  wave
vector  $q_{0}$\cite{KT89,WWSW02,LKBLSS04,ZW05}. As discussed in  section~\ref{sec:textbf-avoid-crit},
these systems all show the  property of avoided criticality.  However,
most of the  studied models are akin  to the Ising version (\textit{i.
e.}, with a discrete $Z_{2}$ symmetry) and  display the "weak" form of
avoided criticality: a fluctuation-induced first-order transition.  We
have  already  mentioned   that this   transition, whose  location  is
non-universal and model-specific,  is hard to  bypass by supercooling,
which  limits the   accessible   domain of  temperature   in  computer
simulation studies. As a result, the conclusions concerning the glassy
behavior of the models may not  be as clear-cut as  one would like and
details concerning the  parameter range studied may  become important:
this may explain the  apparently conflicting results obtained  for the
3-dimensional  Brazovskii-like  model\cite{GR03,GKTV02,SWW03,SW201}.   It
seems anyhow that models with  a continuous symmetry, like the Coulomb
frustrated  $O(N)$  model,  also follow  the  same  scenario  with the
emergence of a complex free-energy landscape below the (truly) avoided
critical temperature $T_{c}^{(0)}$ \cite{K05}.

One may wonder at  this point how reliable  are calculations  based on
approximations which are to a large  extent uncontrolled? There is of
course no definite answer to  this question, but  one can at least try
to check the  robustness of the  scenario: what happens when improved,
or on the contrary worse, approximations are used?  The approximations
being mean-field in character, they may overestimate,  as is often the
case, the presence of transitions and  singularities: are there models
for which the approximations do \textit{not} predict the appearance of
a  complex free-energy landscape?    The available answers to  those
questions tend to support the validity of the scenario. An exponential
number of  metastable   states   is  also  predicted  when   using  an
approximation developed for studying strongly interacting systems with
predominantly local  correlations\cite{WSKW04}, the  "dynamic mean-field
approximation"\cite{GLKR96}; on the other hand, a number of perturbative treatments
cruder than the SCSA mentioned above do not find a complex free-energy
landscape\cite{GKTV02}. Perhaps  more   interestingly, when  the replica
mean-field  approach  and the SCSA are   applied  to the 3-dimensional
model in  the absence of  frustration,  it does \textit{not}  find the
signature of  an     exponentially   large  number    of    metastable
states\cite{M95}, as indeed expected for a pure $\phi^{4}$ theory.

To conclude this  section, it is worth  summarizing the main points of
the  random    first-order   transition    theory    of   glassforming
liquids\cite{KTW89,XW00,XW01,LW01,LW03,MP97}. In this
approach, the  mean-field   scenario  outlined above  is  modified  to
include thermally activated escape from metastable states. The dynamic
mode-coupling-like singularity     is then smeared     out to become a
crossover   to an  activation   dominated regime  associated with  the
evolution of the system on the  complex free-energy landscape. In real
space, the  liquid is described  as  a mosaic  of cells and activation
events as   entropic droplets, the  driving  force for  nuleating such
droplets  being provided   by  the  non-zero configurational  entropy.
Super-Arrhenius temperature  dependence of the $\alpha$-relaxation  time in
the form of a Vogel-Fulcher  expression is predicted as one approaches
the ideal glass transition $T_{K}$; at  this point the configurational
entropy vanishes  and it  can  therefore be  identified to the Kauzman
temperature\cite{K48}. This    approach   successfully describes  many
aspects of the phenomenology of supercooled liquids.

\subsection{Defects and effective kinetic constraints}\label{sec:defects-effect-kinet} 

Topological   defects play   an  important  role  in frustration-based
approaches of glassforming liquids.   In the description of frustrated
icosahedral order   in $3$-dimensional  space for instance,   the most
relevant  defects are  disclinations, \textit{i. e.},  rotational line
defects (see  section~\ref{sec:geom-frustr-simple}).   We  have   also
illustrated in  section~\ref{sec:stat-mech-frustr} how the statistical
mechanical treatment of  frustrated systems may  become more tractable
by  going from  the original  representation in  terms of  local order
parameter to a dual description in terms of topological defects.

It must be emphasized, however, that the simplicity one could gain via
such a transformation  for  studying the structural and  thermodynamic
properties may have a penalty associated  with an increased complexity
in the  effective equation of motion  for the defects. In general, one
would have to deal with  the kinetics of  defect lines in a  partially
ordered but non-crystalline medium. The slow dynamics of defect lines,
such  as  vortex   lines     in  $3$-dimensional  extreme      type-II
superconductors, is  controlled by crossing, cutting and reconnection,
all   processes that  are  thermally   activated and may    lead to an
effective  freezing in an entangled  "glassy" state on the experimental
time  scale\cite{CD95}.    Entanglement  effects are  expected   to be
especially severe for disclination lines  in a frustrated  icosahedral
medium: in    such a case, the  underlying   theory has  a non-Abelian
$SO(4)$ symmetry and the  resulting  combination rules for  the defect
(disclination) lines has been predicted to lead to non-trivial kinetic
constraints   \cite{N83b,N02}.  The phenomenological scaling  approach
reviewed in  section~\ref{sec:phen-scal-appr} side   steps the  issue by merely  considering
quasi-ideally ordered regions (domains) with  a low density of defects
separated  by regions (interfaces) with   a  high density of  defects,
without specifying the nature of the defects nor their kinetics.

To  shed light  on the  subtle   aspects of  the  duality between  the
description in  terms of  local order parameter  and that  in terms of
defects, it is  interesting to  consider  simple models  which do  not
belong   to   the  class  of      frustrated  systems  discussed    in
section~\ref{sec:avoided-criticality}.   They consist  in  Ising  spin
models  on  $2$-dimensional lattices,  with short-ranged ferromagnetic
multi-spin, "plaquette"-type,      interactions.    In   the  original
formulation involving  the   spin variables, the  system  is  strongly
correlated\cite{GN00,G02}   (with  trivial    pair  correlations   but
non-trivial higher-order ones),  and   the dynamics is  described  via
local kinetic rules involving $1$-spin flips.  In  the dual version in
terms of topological defects, the thermodynamics reduces to that of an
ideal gas, but  the effective dynamics of the  defects is now strongly
constrained
\cite{GN00,G02,JBG05}).

The  presence   of  kinetic   constraints,  even in  the    absence of
non-trivial static properties, has been shown to induce a slowing down
of the relaxation and, as a result, formation of glassy states on some
observation time scale.   There is an  abundant amount of work on such
"kinetically constrained models"  (for a review, see Ref.~\cite{RS03}),
which        have   been,        most      notably     in       recent
years\cite{WBG04,GD02,GD03}, advocated  as an  alternative   to thermodynamic
approaches for explaining the phenomenology of supercooled liquids. In
such descriptions,  glassy behavior is  attributed to the emergence of
dynamic heterogeneities,  with a  typical  length scale that grows  in
space-time, but not in space, when temperature is decreased or density
increased. the models reproduce  many features of glassforming liquids,
with  however some shortcomings.  In all  studies  so far carried out,
the defects are   point-like objects; defects  becoming  more and more
dilute as  one      approaches the glass  transition   region,   their
contribution to the entropy and heat capacity of the liquid (in excess
to  those  of  the crystal)   is then completely   negligible, and the
empirical  correlation  between  slowing down  of  the relaxation  and
decrease of the "configurational" (excess) entropy observed in liquids
approaching their glass transition cannot be described
\cite{LW04,BBT04}. In addition,   the  origin  of  the
kinetic constraints is quite elusive.

It is  then tempting  to speculate,  after Palmer  et al.\cite{PSAA84}
that "it is possible to see static frustration over many length scales
as  the underlying  cause of  dynamical   constraints over  many  time
scales". Defects could be   line-like  objects instead  of  point-like
ones,   which could help  keep   the connection  between dynamics  and
thermodynamics.   However, no such   description has been attempted so
far.

\section{Conclusion}\label{sec:conclusion}

In this article,  we  have reviewed the frustration-based  approach of
supercooled liquids and  the glass transition. Frustration in liquids,
namely a ubiquitous incompatibility  between the spatial  extension of
the locally  preferred  liquid structure and  the tiling  of the whole
space, provides a physical mechanism to generate what we have argued to
be the distinctive characteristic of  the viscous slowing down leading
to glass formation: the emergence below  some crossover temperature of
a cooperative behavior whose extension,  however, remains limited to a
mesoscopic   length  scale. Cooperativity    is  associated with   the
(collective) extension of  the locally preferred structure whereas the
limitation   is a    result  of    frustration. Glassiness is     then
self-generated as one decreases temperature.

Despite the absence of a well established minimal theoretical model of
frustration  in glassforming liquids, progress has  been made over the
last decades in  developing a statistical mechanical framework. Models
have  been built, based  on such ideas  as a coupling to a non-Abelian
background, a uniform frustration,  or a competition between effective
interactions. Interestingly,   several generic properties characterize
such  frustrated  models,   largely  independently   of their detailed
structure. Those properties are:  (1)  avoided criticality, a  genuine
non-perturbative phenomenon induced  by frustration, (2) the emergence
of a complex free-energy landscape with  an exponentially large number
of "long-lived" metastable states, and (3) the presence of topological
defects.

The existence of  the  above three  features open  the way to  several
theoretical routes to    explain the  phenomenology   of  glassforming
liquids within a   frustration-based  description. The presence of   a
narrowly avoided critical point in the diagram frustration-temperature
has been used  to  develop  a  phenomenological scaling  analysis   of
supercooled  liquids,  the   frustration-limited  domain  theory.  The
temperature   associated  with the  avoided   critical  point marks  a
crossover  between    "ordinary"   liquid   behavior  and  "anomalous"
(cooperative,  heterogeneous, super-Arrhenius)  liquid   behavior, and
fragility is predicted to be inversely related to frustration. Along a
different line of thought,    the emergence of a  complex  free-energy
landscape is at the basis  of the random first-order transition theory
in which (super-Arrhenius) activated relaxation is associated with the
configurational entropy  resulting from the exponentially large number
of metastable states.   Finally, the existence of topological  defects
(primarily, line  defects   in 3-dimensional  geometrically  frustrated
media),  which can form an  entangled state and  whose dynamics may be
strongly constrained,  is suggestive of a  possible treatment by means
of generalized kinetically  constrained models.  Whether or not  these
theoretical frameworks are compatible  and which of  them is  the most
efficient  to    describe the  physics  of  glass   formation are open
questions.

The phenomenological   scaling  approach about  the  putative  avoided
critical point leads to a good qualitative and even, at the expense of
course of  introducing adjustable parameters, quantitative description
of  the    salient phenomena in    glassforming  liquids. However, its
theoretical foundations, besides  the property of avoided criticality,
remain  shaky. Support  has  been provided   by  computer simulation
studies  of   Coulomb frustrated models,    but no definite conclusion
should   be made.  Needless to  say  that  in  spite  of its appealing
features and of  the results derived so  far, much still remains to be
done to make the frustration-based approach  of supercooled liquids an
operational,  predictive, and microscopically  founded theory of glass
formation.

\section{Acknowledgments}
This work has been  funded by the CNRS and  the NSF. We would  like to
dedicate this article  to Daniel  Kivelson whose stimulating  presence
and creative thinking we miss very much.

%
%\begin{thebibliography}{99}
%

%\end{thebibliography}
%

%

%\bibliography{fld}

\section{References}
\begin{thebibliography}{100}

\bibitem{EAN96}
M.~D. Ediger, C.~A. Angell, and S.~R. Nagel, J. Phys. Chem. {\bf 100},  13200
  (1996).

\bibitem{DEBE01}
P.~G. Debenedetti and F.~H. Stillinger, Nature {\bf 410},  259  (2001).

\bibitem{TARKI01}
G. Tarjus and D. Kivelson,  in {\em Jamming and Rheologiy: Constrained dynamics
  on microscopic and macroscopic scales}, edited by A.~J. Liu and S. Nagel
  (Taylor and Francis, London, 2001).

\bibitem{KTZK96}
D. Kivelson, G. Tarjus, X. Zhao, and S.~A. Kivelson, Phys. Rev. E {\bf 53},
  751  (1996).

\bibitem{CAA85}
C.~A. Angell,  in {\em Relaxation in complex systems}, edited by K.~L. Ngai and
  G.~B. Wright (Office of Naval Research, Arlington, 1985).

\bibitem{A05}
C. Alba-Simionesco, private communication (unpublished).

\bibitem{Silles99}
H. Sillescu, J. Non-Cryst. Solids {\bf 243},  81  (1999).

\bibitem{Edig00}
M.~D. Ediger, Annu. Rev. Phys. Chem. {\bf 51},  99  (2000).

\bibitem{R02}
R. Richert, J. Phys. Cond. Matter {\bf 14},  R703  (2002).

\bibitem{A97}
C.~A. Angell, J. Res. Natl. Inst. Stand. Technol. {\bf 102},  171  (1997).

\bibitem{K48}
W. Kauzmann, Chem. Rev. {\bf 43},  219  (1948).

\bibitem{gotze91}
W. G{\"o}tze,  in {\em Liquids, Freezing and the Glass transition}, edited by
  J.~P.~H. D.~Levesque and J. Zinn-Justin (North Holland, Amsterdam, 1991), p.\
  287.

\bibitem{BB04}
G. Biroli and J. Bouchaud, Europhys. Lett {\bf 67},  21  (2004).

\bibitem{GD03}
J.~P. Garrahan and D. Chandler, Proc. Nat. Acad. Sci. USA {\bf 100},  9710
  (2003).

\bibitem{WBG04}
S. Whitelam, L. Berthier, and J.~P. Garraham, Phys. Rev. Lett. {\bf 92},
  185705  (2004).

\bibitem{AG65}
G. Adams and J.~H. Gibbs, J. Chem. Phys. {\bf 43},  139  (1965).

\bibitem{KTW89}
T.~R. Kirkpatrick, D. Thirumalai, and P.~G. Wolynes, Phys. Rev. A {\bf 40},
  1045  (1989).

\bibitem{XW00}
X. Xia and P.~G. Wolynes, Proc. Nat Acad. Sci USA {\bf 97},  2990  (2000).

\bibitem{MP97}
M. M{\'e}zard and G. Parisi, Phys. Rev. Lett. {\bf 79},  2486  (1997).

\bibitem{KKNT95}
D. Kivelson, S.~A. Kivelson, X.~L. Zhao, Z. Nussinov, and G. Tarjus, Physica A
  {\bf 219},  27  (1995).

\bibitem{SSH91}
J.~P. Sethna, J.~D. Shore, and M. Huang, Phys. Rev. B {\bf 44},  4943  (1991).

\bibitem{ON03}
C.~S. O'Hern, L.~E. Silbert, A.~J. Liu, and S.~R. Nagel, Phys. Rev. E {\bf 68},
   011306  (2003).

\bibitem{G77}
G. Toulouse, Commun. Phys {\bf 2},  115  (1977).

\bibitem{Y98}
{\em Spin Glasses and Random Fields}, edited by A.~P. Young (World Scientific,
  Singapore, 1998).

\bibitem{F52}
F.~C. Frank, Proc. R. Soc. London, Ser. A {\bf 215},  43  (1952).

\bibitem{N02}
D. Nelson, {\em Defects and Geometry in Condensed Matter} (Cambridge University
  Press, Cambridge, 2002).

\bibitem{SM99}
J.~F. Sadoc and R. Mosseri, {\em Geometrical frustration} (Cambridge Univ.
  Press, Cambridge, 1999).

\bibitem{J98}
E.~A. Jagla, Phys. Rev. E {\bf 58},  4701  (1997).

\bibitem{KS79}
M. Kleman and J.~F. Sadoc, J. Phys. Lett. (Paris) {\bf 40},  L569  (1979).

\bibitem{C91}
H.~S.~M. Coxeter, {\em Regular Complex Polytopes} (Cambridge University Press;
  2nd edition, Cambridge, 1991).

\bibitem{MS90}
R. Mosseri and J.~F. Sadoc,  in {\em Geometry in Condensed Matter Physics},
  edited by J.~F. Sadoc (World Scientific, Singapore, 1990),
  Chap.~Non-crystalline atomic structures: from glasses to quasicrystals.

\bibitem{Sa83}
J.~F. Sadoc, J. Phys. lett. {\bf 44},  L  (1983).

\bibitem{SM84}
J.~F. Sadoc and R. Mosseri, J. Physique {\bf 45},  1025  (1984).

\bibitem{N83b}
D.~R. Nelson, Phys. Rev. Lett. {\bf 50},  982  (1983).

\bibitem{N83}
D.~R. Nelson, Phys. Rev. B {\bf 28},  5515  (1983).

\bibitem{NW84}
D.~R. Nelson and M. Widom, Nucl. Phys B {\bf 240},  113  (1984).

\bibitem{SN84}
S. Sachdev and D.~R. Nelson, Phys. Rev. Lett. {\bf 53},  1947  (1984).

\bibitem{SN85}
S. Sachdev and D.~R. Nelson, Phys. Rev. B {\bf 32},  1480  (1985).

\bibitem{Se83}
J.~P. Sethna, Phys. Rev. Lett. {\bf 51},  2198  (1983).

\bibitem{Se85}
J.~P. Sethna, Phys. Rev. B {\bf 31},  6278  (1985).

\bibitem{VD85}
G. Verkataraman and D. Sakoo, Contemp. Phys. {\bf 26},  579  (1985).

\bibitem{VD86}
G. Verkataraman and D. Sakoo, Contemp. Phys. {\bf 27},  3  (1986).

\bibitem{K89}
H. Kleinert, {\em Gauge Fields in Condensed Matter} (World Scientific,
  Singapore, 1989).

\bibitem{Note5}
There is some flexibility in defining the disclinations in the present case.
  Defects are indeed always considered relative to a given, usually ordered,
  ground state. Here, the ground state has perfect icosahedral order but lives
  in a curved space. Disclinations can then either be considered relative to
  the ideal icosahedral template, and the geometric constraints then force an
  excess of disclinations of a given sign (essentially, wedge disclinations
  carrying a 'charge' of $-72^{\circ}$) \cite{N83,Sa83,SM99}; or they can be
  treated as 'curvature carrying' lines, constructed from some ground state
  living on a 'corrugated' space with regions of positive and negative
  curvature but with zero mean curvature: in this case, positive and negative
  disclinations must balance each other to ensure that space is 'flat' on
  average\cite{SM99,SM84}.

\bibitem{FK58}
F.~C. Frank and J.~S. Kasper, Acta Crytallogr. {\bf 11},  184  (1958).

\bibitem{FK59}
F.~C. Frank and J.~S. Kasper, Acta Crytallogr. {\bf 12},  483  (1958).

\bibitem{FFMMNRSST03}
F. Faupel, W. Frank, M.~P. Macht, H. Mehrer, V. Naundorf, K. R{\"a}tzke, H.~R.
  Schoder, S.~K. Sharma, and H. Teichler, Rev. Mod. Phys. {\bf 75},  237
  (2003).

\bibitem{TKV00}
G. Tarjus, D. Kivelson, and P. Viot, J. Phys: Cond. Mat. {\bf 12},  6497
  (2000).

\bibitem{MT03}
S. Mossa and G. Tarjus, J. Chem. Phys. {\bf 119},  8069  (2003).

\bibitem{SNR81}
P.~J. Steinhard, D.~R. Nelson, and M. Ronchetti, Phys. Rev. Lett. {\bf 47},
  1297  (1981).

\bibitem{SNR83}
P.~J. Steinhard, D.~R. Nelson, and M. Ronchetti, Phys. Rev. B {\bf 28},  784
  (1983).

\bibitem{S92}
{\em Bond-Orientational Order in Condensed-Matter Systems}, edited by K.~J.
  Standburg (Springer-Verlag, Berlin, 1992).

\bibitem{YABW96}
Y.~L. Yarger, C.~A. Angell, S.~S. Borick, and G.~H. Wolf,  in {\em Supercooled
  Liquids: Advances and Novel Applications}, edited by J.~T.~F. et.al (ACS
  Symposium Series (American Chemical Society, Whashington DC, 1996), Vol.~676,
  p.\ 214.

\bibitem{MS98}
O. Mishima and H.~E. Stanley, Nature {\bf 396},  329  (1998).

\bibitem{TAGVK03}
G. Tarjus, C. Alba-Simionesco, M. Grousson, P. Viot, and D. Kivelson, J. Phys:
  Cond. Matter {\bf 15},  S1077  (2003).

\bibitem{NT81}
D.~R. Nelson and J. Toner, Phys. Rev. B {\bf 24},  363  (1981).

\bibitem{S84}
J.~P. Straley, Phys. Rev. B {\bf 30},  6592  (1984).

\bibitem{Note8}
Stated otherwise, one has first to look for a minimal theoretical model which
  could play a role similar to the Edwards-Anderson model for the description
  of spin glasses\cite{EA75}. The situation however is far from being as well
  founded as in the case of spin glasses for which the known key-ingredients,
  quenched disorder and frustration, are appropriately accounted for by the
  Edwards-Anderson model.

\bibitem{N04}
Z. Nussinov, Phys. Rev. B {\bf 69},  014208  (2004).

\bibitem{CafLB86}
R.~G. Caflish, H. Levine, and J. Banavar, Phys. Rev. Lett. {\bf 57},  2679
  (1986).

\bibitem{LL60}
L.~D. Landau and E.~M. Lifschitz, {\em The classical theory of fields, 4th ed.}
  (Pergamon Press, Oxford England, 1984).

\bibitem{M79}
D. Mermin, Rev. Mod. Phys. {\bf 51},  591  (1979).

\bibitem{RD82}
N. Rivier and D.~M. Dufty, J. Phys. (Paris) {\bf 43},  293  (1982).

\bibitem{R83}
N. Rivier,  in {\em Topological Disorder in Condensed Matter}, edited by F.
  Yonezawa and T. Ninomiya (Springer Verlag, Berlin, 1983), p.\ 13.

\bibitem{K93}
I. Kanazawa, Phys. Lett. A {\bf 176},  246  (1993).

\bibitem{K02}
I. Kanazawa, J. Non-Cryst. Solids {\bf 293-295},  615  (1993).

\bibitem{T80}
M. Tinkham, {\em Introduction to Superconductivity, 2nd Edition} (Mc Graw-Hill,
  New York, 1996).

\bibitem{RN83}
M. Rubinstein and D.~R. Nelson, Phys. Rev. B {\bf 28},  6377  (1983).

\bibitem{PL96}
J.~M. Park and T.~C. Lubensky, Phys. Rev. E {\bf 53},  2648  (1996).

\bibitem{BNT00}
M. Bowick, D.~R. Nelson, and A. Travesset, Phys. Rev. B {\bf 62},  8738
  (2000).

\bibitem{FT94}
M. Franz and S. Teitel, Phys. Rev. Lett. {\bf 73},  480  (1994).

\bibitem{FT95}
M. Franz and S. Teitel, Phys. Rev. B {\bf 51},  6551  (1995).

\bibitem{F80}
D. Fisher, Phys. Rev. B {\bf 22},  1190  (1980).

\bibitem{Note0}
We focus here on models in which frustration can be continuously tuned. We do
  not consider uniformly frustrated spin models where frustration results from
  the geopmetry of the lattice, such the already mentioned antiferromagnetic
  Ising model on a triangular lattice. Such systems usually display a large
  degeneracy of the ground state, a complex phase diagram, and other
  interesting phenomena \cite{D04}, but we do not think they are directly
  relevant to the glass transition problem.

\bibitem{JKKN77}
J.~V. Jos{\'e}, L.~P. Kadanoff, S. Kirkpatrick, and D.~R. Nelson, Phys. Rev. B
  {\bf 16},  1217  (1977).

\bibitem{KL97}
B. Kim and S.~J. Lee, Phys. Rev. Lett. {\bf 78},  3709  (1997).

\bibitem{LK99}
S.~J. Lee and B. Kim, Phys. Rev. E {\bf 60},  1503  (1999).

\bibitem{LT93}
Y.~H. Li and S. Teitel, Phys. Rev. B {\bf 47},  359  (1993).

\bibitem{Note1}
Notice that the gradient term expressed with the new variable
  $\mathbf{\tilde{Q}}(\mathbf{x})$ (see Eq.(\ref{eq:21}) is quite similar to
  that derived for the 2-dimensional uniformly frustrated systems, Eqs.
  (\ref{eq:7}) and (\ref{eq:9}). In all cases there is a connection that is
  akin to a fixed, but now non-uniform gauge field. The gauge group however is
  Abelian for superconducting films ( $U(1)$), whereas it is non-Abelian in
  theory of geometric frustration (e. g., $SO(4)$). Topological defects are
  induced by frustration in all cases, but the consequences of the Abelian
  versus non-Abelian property must still be investigated.

\bibitem{GSM82}
J.~P. Gaspard, J.~F. Sadoc, and R. Mosseri,  in {\em The Structure of
  Non-Crystalline Materials}, edited by P. Gaskell, X. Parker, and Y. Davis
  (Taylor and Francis, London, 1982), p.\ 550.

\bibitem{GKT97}
Y.~N. Gornostyrev, M.~I. Katsnelson, and A.~V. Trefilov, J. Phys: Condens.
  Matter {\bf 9},  7837  (1997).

\bibitem{L80}
L. Leibler, Macromolecules {\bf 13},  1602  (1980).

\bibitem{OK86}
T. Ohta and K. Kawazaki, Macromolecules {\bf 19},  2621  (1986).

\bibitem{BF90}
F. Bates and G. Fredrickson, Ann. Rev. Phys. Chem. {\bf 41},  525  (1990).

\bibitem{WCS92}
D. Wu, D. Chandler, and B. Smit, J. Phys. Chem. {\bf 96},  4077  (1992).

\bibitem{DC94}
M.~W. Deem and D. Chandler, Phys. Rev. E {\bf 49},  4268  (1994).

\bibitem{WWSW02}
S. Wu, H. Westfahl, J. Schmalian, and P.~G. Wolynes, Chem. Phys. Lett. {\bf
  359},  1  (2002).

\bibitem{S83}
F.~H. Stillinger, J. Chem. Phys. {\bf 78},  4654  (1983).

\bibitem{EK93}
V.~J. Emery and S.~A. Kivelson, Physica C {\bf 209},  597  (1993).

\bibitem{CEKNT96}
L. Chayes, V.~J. Emery, S.~A. Kivelson, Z. Nussinov, and G. Tarjus, Physica A
  {\bf 225},  129  (1996).

\bibitem{NRKC99}
Z. Nussinov, J. Rudnick, S.~A. Kivelson, and L.~N. Chayes, Phys. Rev. Lett.
  {\bf 83},  472  (1999).

\bibitem{JKS05}
R. Jamei, S.~A. Kivelson, and B. Spivak, Phys. Rev. Lett. {\bf 94},  056805
  (2005).

\bibitem{B75}
S.~A. Brazovskii, Sov. Phys. JETP {\bf 41},  85  (1975).

\bibitem{VT98}
P. Viot and G. Tarjus, Europhys. Lett {\bf 44},  423  (1998).

\bibitem{GTV01b}
M. Grousson, G. Tarjus, and P. Viot, Phys. Rev. E {\bf 64},  036109  (2001).

\bibitem{N01}
Z. Nussinov, Cond-mat {\bf 0105253},    (2001).

\bibitem{Nelson83}
D.~R. Nelson,  in {\em Phase Transitions and Critical Phenomena}, edited by C.
  Domb and J. Lebowitz (Academic Press, New York, 1983), p.\ 1.

\bibitem{KT72}
J.~M. Kosterlitz and D.~J. Thouless, J. Phys. C {\bf 5},  L121  (1972).

\bibitem{V72}
V.~L. Berezinskii, Sov. Phys. JETP {\bf 34},  610  (1972).

\bibitem{GT00}
V. Gotcheva and S. Teitel, Phys. Rev. Lett. {\bf 86},  2126  (2000).

\bibitem{T99}
Z. Tesanovic, Phys. Rev. B {\bf 59},  6449  (1999).

\bibitem{OT05}
S. Teitel and P. Olson, Phys. Rev. Lett. {\bf 94},  219703  (2005).

\bibitem{AP83}
A. Aharony and E. Pytte, Phys. Rev. B {\bf 27},  5872  (1983).

\bibitem{KZKFK94}
S.~A. Kivelson, X. Zhao, D. Kivelson, T.~M. Fischer, and C.~M. Knobler, J.
  Chem. Phys. {\bf 101},  2391  (1994).

\bibitem{Note2}
One of the weaknesses of the scaling approach so far developed for
  frustration-induced avoided criticality is the elusiveness concerning the
  nature of the order variable and its coupling to the fluctuations of the main
  observables such as the one-body density or the dipole density. Some of the
  assumptions, e. g., $\theta=2$, are easily justified if the relevant variable
  has a discrete symmetry, but as discussed in section
  \ref{sec:stat-mech-frustr} this discreteness is by no means obvious.

\bibitem{FHS88}
F.~H. Stillinger, J. Chem. Phys. {\bf 89},  6461  (1988).

\bibitem{VTK00}
P. Viot, G. Tarjus, and D. Kivelson, J. Chem. Phys. {\bf 112},  10368  (2000).

\bibitem{ALBA01}
C. Alba-Simionesco, C. R. A. S. {\bf IV},  1114  (2001).

\bibitem{DWNWC90}
P. Dixon, L. Wu, S.~R. Nagel, B.~D. Williams, and J.~P. Carini, Phys. Rev.
  Lett. {\bf 65},  1108  (1990).

\bibitem{TK95}
G. Tarjus and D. Kivelson, J. Chem. Phys. {\bf 103},  3071  (1995).

\bibitem{RS03}
F. Ritort and P. Sollich, Adv. Phys. {\bf 52},  219  (2003).

\bibitem{HF87}
D.~A. Huse and D.~S. Fisher, Phys. Rev. B {\bf 35},  6841  (1987).

\bibitem{KT98}
D. Kivelson and G. Tarjus, J. Chem. Phys. {\bf 109},  5481  (1998).

\bibitem{GTV02}
M. Grousson, G. Tarjus, and P. Viot, J. Phys.: Cond. Mat. {\bf 14},  1617
  (2002).

\bibitem{GTV02b}
M. Grousson, G. Tarjus, and P. Viot, Phys. Rev. E {\bf 65},  065103(R)  (2002).

\bibitem{GTV01}
M. Grousson, G. Tarjus, and P. Viot, Phys. Rev. Lett. {\bf 86},  3455  (2001).

\bibitem{D04}
{\em Frustrated spin systems}, edited by H.~T. Diep (World Scientific,
  Singapore, 2004).

\bibitem{BK01}
G. Biroli and J. Kurchan, Phys. Rev. E {\bf 64},  106101  (2001).

\bibitem{M95}
R. Monasson, Phys. Rev. Lett. {\bf 75},  2875  (1995).

\bibitem{Note6}
For a pedagogical introduction, see T. Castellini and A. Cavagna,
  cond-mat/0505032.

\bibitem{ScWo00}
J. Schmalian and P. Wolynes, Phys. Rev. Lett. {\bf 85},  836  (2000).

\bibitem{WSW01}
H. Westfahl, J. Schmalian, and P. Wolynes, Phys. Rev. B {\bf 64},  174203
  (2001).

\bibitem{AJBa74}
A.~J. Bray, Phys. Rev. Lett. {\bf 32},  1413  (1974).

\bibitem{Note4}
The SCSA amounts to formally consider an $O(N)$ version of the model under
  study in the limit of large number of components $N$, perform a perturbative
  expansion in $1/N$ and include in a self-consistent manner the leading $1/N$
  corrections; one then obtains a coupled set of self-consistent equations for
  the pair correlation functions of the model. $N$ is then set to the chosen
  value, in most case $N=1$. The same type of approximation can be derived for
  the dynamics of the system\cite{BCKM96,B02}.

\bibitem{GKTV02}
M. Grousson, V. Krakoviack, G. Tarjus, and P. Viot, Phys. Rev. E {\bf 66},
  026126  (2002).

\bibitem{XW01}
X. Xia and P.~G. Wolynes, Phys. Rev. Lett. {\bf 86},  5526  (2001).

\bibitem{LW01}
V. Lubchenko and P.~G. Wolynes, Phys. Rev. Lett. {\bf 87},  195901  (2001).

\bibitem{LW03}
V. Lubchenko and P.~G. Wolynes, J. Chem. Phys. {\bf 119},  9088  (2003).

\bibitem{Note7}
Note that the ground-state entropy in these systems (at least in their large
  $N$ behavior) can be rigourously shown to be proportional to the surface area
  of the system. The extremely high ground-state degeneracy in these frustrated
  systems is initimately linked to the existence of a large number of
  unconventional symmetries\cite{N04,N01,BN05}. Yet the entropy density goes to
  zero in the thermodynamic limit.

\bibitem{KT89}
T.~R. Kirkpatrick and D. Thirumalai, J. Phys. A: Math. Gen. {\bf 22},  L149
  (1989).

\bibitem{LKBLSS04}
K.~K. Loh, K. Kawasaki, A.~R. Bishop, T. Lookman, A. Saxena, Z. Nussinov, and
  J. Schmalian, Phys. Rev. E {\bf 69},  010501  (2004).

\bibitem{ZW05}
C.~Z. Zhang and Z.~G. Wang, Cond-mat {\bf 053536},    (2005).

\bibitem{GR03}
P.~L. Geissler and D.~R. Reichman, Phys. Rev. E {\bf 69},  021501  (2003).

\bibitem{SWW03}
J. Schmalian, P.~G. Wolynes, and S. Wu, Cond-mat {\bf 0305420},    (2003).

\bibitem{SW201}
J. Schmalian and P. Wolynes, Phys. Rev. Lett. {\bf 86},  3456  (2001).

\bibitem{K05}
V. Krakoviack, private communication (unpublished).

\bibitem{WSKW04}
S. Wu, J. Schmalian, G. Kotliar, and P.~G. Wolynes, Phys. Rev. B {\bf 70},
  024207  (2004).

\bibitem{GLKR96}
A. Georges, G. Kotliar, W. Krauth, and M. Rozenberg, Rev. Mod. Phys. {\bf 68},
  13  (1996).

\bibitem{CD95}
C. Carraro and D.~S. Fisher, Phys. Rev. B {\bf 51},  534  (1995).

\bibitem{GN00}
J.~P. Garrahan and M.~E.~J. Newmann, Phys. Rev. E {\bf 62},  7670  (2000).

\bibitem{G02}
J.~P. Garrahan, J. Phys: Condens. Matter {\bf 14},  1571  (2002).

\bibitem{JBG05}
R.~L. Jack, L. Berthier, and J.~P. Garrahan, Cond-mat {\bf 0502120},    (2005).

\bibitem{GD02}
J.~P. Garrahan and D. Chandler, Phys. Rev. Lett. {\bf 89},  035704  (2002).

\bibitem{LW04}
V. Lubchenko and P.~G. Wolynes, J. Chem. Phys. {\bf 121},  2852  (2004).

\bibitem{BBT04}
G. Biroli, J.-P. Bouchaud, and G. Tarjus, Cond-mat {\bf 0412024},    (2004).

\bibitem{PSAA84}
R.~G. Palmer, D.~L. Stein, E. Abrahams, and P.~W. Anderson, Phys. Rev. Lett.
  {\bf 53},  958  (1984).

\bibitem{EA75}
S.~F. Edwards and P.~W. Anderson, J. Phys. F: Metal Phys. {\bf 5},  965
  (1975).

\bibitem{BCKM96}
J.~P. Bouchaud, L. Cugliandolo, J. Kurchan, and M. M{\'e}zard, Physica A {\bf
  226},  243  (1996).

\bibitem{B02}
J. Berges, Nucl. Phys. A {\bf 699},  847  (2002).

\bibitem{BN05}
C. Batista and Z. Nussinov, Cond-mat {\bf 0410599},    (2004).

\end{thebibliography}

\end{document}